\documentclass[preprintnumbers,amsmath,amssymb,floatfix,10pt,prd,onecolumn,
superscriptaddress,nofootinbib]{revtex4}
\usepackage{latexsym}
\usepackage{epsfig}
\usepackage{epstopdf}
\usepackage{graphicx}
\usepackage{amssymb}
\usepackage{amsmath}
\usepackage{dcolumn}
\usepackage{bm}
\usepackage{color}
\usepackage{comment}
\usepackage[utf8]{inputenc}
\begin{document}

\title{\bf Thermodynamics and Perturbative Analysis of Some Newly Developed $\mathcal{F}(R,L_m, T)$ Theories
Under the Scenario of Conserved Energy-momentum Tensor}

\author{M. Zubair}
\email{mzubairkk@gmail.com; drmzubair@cuilahore.edu.pk}\affiliation{Department
of Mathematics, COMSATS University Islamabad, Lahore Campus, Lahore,
Pakistan}
\author{Saira Waheed}
\email{swaheed@pmu.edu.sa}\affiliation{Prince Mohammad Bin Fahd University, Al Khobar, 31952 Kingdom of Saudi Arabia}
\author{Quratulien Muneer}
\email{anie.muneer@gmail.com}\affiliation{Department of Mathematics, COMSATS University Islamabad, Lahore Campus, Lahore, Pakistan}
\author{M. Ahmad}
\email{ahmadhunny786@gmail.com}\affiliation{Department
of Mathematics, COMSATS University Islamabad, Lahore Campus, Lahore,
Pakistan}

\begin{abstract}

The present work is devoted to explore some interesting cosmological features of a newly proposed theory of
gravity namely $\mathcal{F}(R,L_m,T)$ theory, where $R$ and $T$ represent the Ricci scalar and trace of energy
momentum-tensor, respectively. Firstly, a non-equilibrium thermodynamical description is considered on the
apparent horizon of the Friedmann's cosmos. The Friedmann equations are demonstrated to be equivalent to
the first law of thermodynamics, i.e., ${T_{Ah}d\varepsilon_{h}^\prime+T_{Ah}d_{i}\varepsilon_{h}^\prime=-d\hat{E}+\hat{W}dV}$,
where ${d_{i}\varepsilon_{h}^\prime}$ refers to entropy production term. We also formulate the constraint for validity of
generalized second law of thermodynamics and check it for some simple well-known forms of generic function $\mathcal{F}(R,L_m,T)$.
Next, we develop the energy bounds for this framework and constraint the free variables by finding
the validity regions for NEC and WEC. Further, we reconstruct some interesting cosmological solutions namely
power law, $\Lambda$CDM and de Sitter models in this theory. The reconstructed solutions are then examined by
checking the validity of GSLT and energy bounds. Lastly, we analyze the stability of all reconstructed
solutions by introducing suitable perturbations in the field equations. It is concluded that obtained
solutions are stable and cosmologically viable.\\\\
\textbf{Keywords:}  $F(R,L_m,T)$ gravity, Thermodynamics, Energy Conditions, Reconstruction.
\end{abstract}

\maketitle

\date{\today}

\section{Introduction}

The formulation of a modified gravity theory which can describe the phenomenon of
current cosmic expansion successfully and provide answers to other cosmic challenges
accurately is regarded as one of the most interesting topics in modern cosmology.
In recent past, modified gravity theories approach has been emerged as a powerful tool where
the modifications in the gravitational part of Einstein-Lagrangian are introduced to accommodate
dark ingredients. In this respect, some well-known examples include scalar-tensor extensions of GR \cite{1m,2m},
torsion based formulations \cite{3m,4m} and Gauss-Bonnet theory and its extensions \cite{5m,6m}, $\mathcal{F}(R)$
gravity and theories involving higher-order curvature corrections \cite{7m}-\cite{9m} and the
gravitational frameworks involving coupling of curvature and matter \cite{10m}-\cite{14m}. It
is argued in literature that the modified gravitational frameworks can provide a successful
explanation to various cosmic issues.

In 1970, Hans Adolph Buchdahl proposed a gravity theory named as $\mathcal{F}(R)$ formalism \cite{15m}
which was a simple extension of Einstein's theory of general relativity (GR) by substituting a generic
function $\mathcal{F}(R)$ instead of $R$. Later, different attempts were made to find a gravitational
framework involving matter and curvature interaction, for example, in \cite{11m,12m}, authors proposed the
action $I=\int\sqrt{-g}(\mathcal{F}_1(R)+(1+\lambda \mathcal{F}_2(R))L_m)dx^4$. In this respect, one significant modificationwas presented by Har ko et al. \cite{13m} where they introduced an arbitrary function of $R$ and $L_{m}$ and is known as $\mathcal{F}(R,L_{m})$ gravity. It is argued that one of the important aspects of this theory is the presence of an extra force which leads to non-geodesic motion in this framework and consequently, some significant differences in results of various cosmological phenomenon can be noticed when compared to the predictions of GR or other extended theories \cite{16m*}. In \cite{16m}, the existence of Neutron star models has been explored within this gravitational framework and obtained results are tested for GW170817 and PSRJ0030+0451 observations. In another work \cite{R1}, the scenario of cosmic expansion is investigated in $\mathcal{F}(R,L_{m})$ framework by taking a non-linear model into account and using Om diagnostic parameter, it is found that their model exhibits quintessence behavior. In recent literature, Harko and Shahidi \cite{R2} discussed matter to geometry interaction in conformal quadratic Weyl gravity by assuming a specific coupling term. In another work \cite{R3}, same authors investigated such interaction by using Palatini formalism and compared their results with the standard $\Lambda$CDM model.\\
Another interesting extended version of $\mathcal{F}(R)$ framework was proposed by Harko et al. \cite{14m}
where they involved contribution from $T$ (trace of energy-momentum tensor (EMT)) and is known as $\mathcal{F}(R,T)$ gravity. They obtained the relevant field equations in metric formalism for some exceptional circumstances involving the connection of matter and curvature. In \cite{R4}, authors explored the equations of motion for massive test particles in the Palatini formulation of $\mathcal{F}(R,T)$ theory and provided a brief thermodynamical interpretation of this framework. In $\mathcal{F}(R,T)$ gravity, researchers have computed various astrophysical and cosmological aspects in detail in \cite{18m,19m, R6, R7,zub1, zub2, zub3}. Further, another important extension namely $\mathcal{F}(R,T^2)$ theory has been proposed in literature \cite{20m,21m} where $T^2=T_{\mu\nu}T^{\mu\nu}$.  Tretyakov \cite{R5} proposed another modification of $\mathcal{F}(R,T)$ theory by taking higher derivatives matter field and discussed stability conditions in the proposed framework for possible avoidance of primary instabilities like ghost and tachyon instabilities. Although, the $\mathcal{F}(R,T)$ theory has developed an intriguing
extension to GR, showing a wide range of phenomenology in astrophysics and cosmology but in \cite{R8}, author addressed the challenges that arise when attempting to explain a workable and realistic cosmology within the $\mathcal{F}(R,T)$ class of theories. They demonstrated that the $\mathcal{F}(R,T)$ models of the form $f(R,T)$ that are currently accessible in the literature do not result in an expandable cosmological background. To overcome such difficulties, Haghani and Harko \cite{22m} expanded, generalized, and simultaneously unified two kinds of gravitational theories with geometry-matter coupling, namely the $\mathcal{F}(R, L_m)$ and the $\mathcal{F}(R, T)$ theories. They took into account a gravitational theory where matter is non-minimally coupled with geometry at the levels of the matter Lagrangian $L_m$ and the trace of the energy-momentum tensor $T$, with the Lagrangian density given by $\mathcal{F}(R, L_m, T)$. One significant feature of such alternative schemes is the involvement of an additional force which arise from the non-geodesic movement. They explored some important cosmological aspects like Poisson equation, Newtonian limit, and the Dolgov-Kawasaki instability in this theory. This study unfolds new directions to explore the theoretical, observational and even experimental aspects of the large class of non-minimally coupled theories.

In this manuscript our main aim is to find the generalized results which will retrieve those obtained in $f(R,T)$, $f(R,L_m)$ and $f(R)$ theories. $f(R,L_m, T)$ theory appeared as more comprehensive as compared to those already developed in literature because here non-minimal coupling of geometry occurs at both the levels of the matter Lagrangian $L_m$, and of the trace of the energy–momentum tensor $T$. As a first step
in our study, we obtain the thermodynamic laws which are tested for various models in $f(R,L_m, T)$ modified theory. In this theory energy–momentum tensor is not conserved anymore, we establish a particular scenario of finding new $f(R,L_m, T)$ models under the scenario of conserved energy–momentum tensor. Stability of newly reconstructed solutions is tested against the homogenous perturbations followed by Dolgov-Kawasaki instability criterion. Moreover limitations are developed on the basis of generalized second law of thermodynamics and energy conditions. The present article is arranged with the following sequence of sections.
Upcoming section will describe the basic formalism of $\mathcal{F}(R,L_m, T)$ gravity and its field equations in detail. In section $\textbf{III}$, the laws of thermodynamics in this framework will be presented and the validity of corresponding constraint will be investigated for some well-known simple $\mathcal{F}(R,L_m,T)$ functions. Section $\textbf{IV}$ will develop the general energy condition bounds for this theory and the possible ranges for free parameters will be explored graphically for the same choices of $\mathcal{F}(R,L_m,T)$ function. In Section $\textbf{V}$,
we shall discuss the reconstruction of power law model and $\Lambda$CDM models and discuss the validity
of generalized second law of thermodynamics (GSLT) as well as energy conditions for the obtained functions. In Section $\textbf{VI}$, we shall investigate the stability of reconstructed solutions by introducing perturbations for all cases. Last section will summarize and conclude the whole study.

\section{Preliminaries of $\mathcal{F}(R,L_{m},T)$ Theory of Gravity}

In this section, we shall briefly describe the basic formulation of $\mathcal{F}(R,L_{m},T)$ gravity
along with its field equations. We shall also provide some essential assumptions used for this work.
The action of $\mathcal{F}(R,L_{m},T)$ gravity is given by \cite{22m}
\begin{eqnarray}\label{1}
S&=&\frac{1}{16\pi}\int{\mathcal{F}(R,L_{m},T)\sqrt{-g} d^{4}x}+\int{L_{m}\sqrt{-g} d^{4}x},
\end{eqnarray}
where $\mathcal{F}(R,L_{m},T)$ is a generic function of Ricci scalar and trace of EMT while
$L_{m}$ denotes the matter Lagrangian density.
The EMT of ordinary matter is defined as \cite{56}
\begin{eqnarray}\label{2}
T_{\alpha\beta}&=&-\frac{2}{\sqrt{-g}}\frac{\delta(\sqrt{-g}L_{m})}{\delta{g^{\alpha\beta}}}.
\end{eqnarray}
In the present work, we assume that $L_{m}$ depends only on the components of metric-tensor $g_{\alpha\beta}$ so that
\begin{eqnarray}\label{3}
T_{\alpha\beta}&=&g_{\alpha\beta}L_{m}-2\frac{\partial{L_{m}}}{\partial{g_{\alpha\beta}}}.
\end{eqnarray}
The EMT of ordinary matter is considered to be the perfect fluid and is given by
\begin{eqnarray}\label{4}
T_{\alpha\beta}^{matter}&=&(\rho+p)u_{\alpha}u_{\beta}+p g_{\alpha\beta},
\end{eqnarray}
where $u_{\alpha}$ is the fluid's four-velocity. The terms $\rho$ and $p$ denote the ordinary matter density and
pressure, respectively. Taking the variation of Eq.(\ref{1}) with respect to
$g_{\alpha\beta}$ yields the following set of field equations:
\begin{eqnarray}\label{5}
\mathcal{F}_{R}R_{\alpha\beta}-\frac{1}{2}[\mathcal{F}-(\mathcal{F}_{L}+2\mathcal{F}_{T})L_{m}]
g_{\alpha\beta}+(g_{\alpha\beta}\Box-\nabla_{\alpha}\nabla_{\beta})\mathcal{F}_{R}
&=&[8\pi G+\frac{1}{2}(\mathcal{F}_L+2\mathcal{F}_T)]T_{\alpha \beta} +\mathcal{F}_{T} \tau_{\alpha\beta},
\end{eqnarray}
where the introduced notations are defined as
$\tau_{\alpha\beta}=2g^{ij}\frac{\partial^2{L_m}}{\partial{g^{ij}}\partial{g^{\alpha\beta}}}$,
$\Box=g^{\alpha \beta}\nabla_{\alpha}\nabla_{\beta}$ while $\nabla_{\alpha}$ represents the covariant derivative.
Also, the terms $\mathcal{F}_{R}$, $\mathcal{F}_{L}$ and $\mathcal{F}_{T}$ stand for the derivative of generic
function with respect to subscript variable. It is important to mention here that the divergence of EMT is not
conserved in this gravity and yields the following relation:
\begin{eqnarray}\label{6}
\bigtriangledown^{\alpha}T_{\alpha \beta}&=&\frac{1}{8\pi G+\mathcal{F}_m}\bigg[\bigtriangledown_{\beta}(L_m \mathcal{F}_m)
-T_{\alpha\beta}\bigtriangledown^{\alpha} \mathcal{F}_m-A_{\beta}-\frac{1}{2}
(\mathcal{F}_T\bigtriangledown_{\beta}T+\mathcal{F}_L \bigtriangledown_{\beta} L_m)\bigg],
\end{eqnarray}
where $A_{\beta}=\bigtriangledown^{\alpha}(\mathcal{F}_T \tau_{\alpha \beta})$ and
$\mathcal{F}_m=\mathcal{F}_T+\frac{1}{2}\mathcal{F}_L$. The standard continuity equation
holds true if $\mathcal{F}_T=\mathcal{F}_L=0$. Also, it is worthy to mention here that $A_\beta=0$
for perfect fluid as well as scalar field theory.

The metric for non-flat FRW cosmos is given by
\begin{eqnarray}\label{7}
 ds^2&=&-dt^2+\frac{a^2}{1-kr^2}dr^2+a^2d\theta^2+a^2sin^2\theta d\phi^2
\end{eqnarray}
with $a(t)$ being a scale factor, while $k$ denotes the curvature of the universe ($k=1,~ 0,~ -1$ refer to
closed, flat and open universes, respectively). For FRW geometry and perfect fluid matter, the $00$-component of
Eq.(\ref{5}) is given by
\begin{eqnarray}\label{8}
&&8\pi G \rho+ \frac{1}{2}(\mathcal{F}_L+2\mathcal{F}_T)(\rho + L_m)-\frac{\mathcal{F}}{2}+3(\dot{H}+H^2)\mathcal{F}_{R}-3H(\dot{R}f_{RR}+\dot{T}\mathcal{F}_{RT}
+\dot{L}_m\mathcal{F}_{R L})=0.
\end{eqnarray}
Here, the dot indicates time derivative while the Hubble parameter, Ricci scalar and trace of EMT can be, respectively,
defined as $H=\frac{\dot{a}}{a},~\mathcal{R}=6(\dot{H}+2 H^{2}+\frac{k}{a^2})$ and $T=-\rho+3p$. Equation (\ref{6}) can be
re-written as follows
\begin{eqnarray}\label{9}
\dot{\rho}+3H(\rho+p)&=&\frac{-1}{8\pi G+\mathcal{F}_m}\bigg[(\rho+L_m)\dot{\mathcal{F}}_m-\frac{1}{2}
\mathcal{F}_{T}(\dot{T}-2\dot{L}_m)\bigg]
\end{eqnarray}
which indicates that $\bigtriangledown^{\alpha}T_{\alpha\beta}\neq 0$ in $\mathcal{F}(R,L_m,T)$ gravity.
In order to obtain the continuity equation in standard form, one needs to impose the condition:
$\bigtriangledown^{\alpha}T_{\alpha\beta}= 0$ which further yields the following constraint to be satisfied
$$\frac{-1}{8\pi G+\mathcal{F}_m}\bigg[(\rho+L_m)\dot{\mathcal{F}}_m-\frac{1}{2}\mathcal{F}_{T}(\dot{T}-2\dot{L}_m)\bigg]=0.$$
Applying the EoS $p=\omega\rho$ in the last two equations, we obtain
\begin{eqnarray}\label{10*}
&&\dot{\rho}+3 H \rho(1+\omega)=0,\\\label{10}
&&(\rho+L_m)[\mathcal{F}_{TT}\dot{T}+\dot{L}\mathcal{F}_{TL}+\dot{R}\mathcal{F}_{TR}
+\frac{\dot{L}}{2}\mathcal{F}_{LL}+\frac{\dot{T}}{2}\mathcal{F}_{LT}+\frac{\dot{R}}{2}\mathcal{F}_{LR}]-
\frac{1}{2}\mathcal{F}_T(\dot{T}-2\dot{L}_m)=0.
\end{eqnarray}
The field equations in the background of FRW universe model can be written as
\begin{eqnarray}\label{11}
&&3(H^2+\frac{k}{a^2})\mathcal{F}_R=8\pi G \rho +\frac{1}{2}(\mathcal{F}_L+2\mathcal{F}_T)(\rho+L_m)
+\frac{1}{2}(R\mathcal{F}_R-\mathcal{F})-3H(\dot{R}\mathcal{F}_{RR}+\dot{T}\mathcal{F}_{RT}+\dot{L}_{m}\mathcal{F}_{RL}),\\\nonumber
&&-(2\dot{H}+3H^2+\frac{k}{a^2})\mathcal{F}_R=8\pi G p+\frac{1}{2}(\mathcal{F}_L+2\mathcal{F}_T)(p-L_m)
+2H(\dot{R}\mathcal{F}_{RR}+\dot{T}\mathcal{F}_{RT}
+\dot{L}_m\mathcal{F}_{RL})+\ddot{R}\mathcal{F}_{RR}+\\\label{12}
&&\dot{R}^{2}\mathcal{F}_{RRR}+2\dot{R}\dot{T}\mathcal{F}_{RRT}-\frac{1}{2}(R\mathcal{F}_R -\mathcal{F})
+\ddot{T}\mathcal{F}_{RT}+\dot{T}^2\mathcal{F}_{RTT}+2\dot{R}\dot{L}_{m}\mathcal{F}_{RRL}
+\ddot{L}_{m}\mathcal{F}_{RL}+\dot{L}_{m}^{2}\mathcal{F}_{RLL}.
\end{eqnarray}
Equation (\ref{5}) can be written as {$G_{\alpha\beta}=8\pi G_{eff}T_{\alpha\beta}^{eff}$}, where
$G_{\alpha\beta}=R_{\alpha\beta}-\frac{1}{2}R g_{\alpha\beta}$, $G_{eff}=\frac{G\hat{Y}}{\mathcal{F}_R}$
and $T^{eff}_{\alpha\beta}=T^{matter}_{\alpha\beta}+T^{dark}_{\alpha\beta}$. Also, $\hat{Y}=1+\frac{\mathcal{F}_m}{8\pi G}$.
Further, the part of EMT involving dark source terms is defined as
\begin{eqnarray}\label{t*}
T^{dark}_{\alpha\beta}&=&\frac{1}{8\pi G+\mathcal{F}_m}\bigg[\frac{g_{\alpha\beta}}{2}(\mathcal{F}-R\mathcal{F}_R)
-g_{\alpha\beta}\mathcal{F}_m L_m+(\nabla_{\alpha}\nabla_{\beta}-g_{\alpha\beta}\Box)\mathcal{F}_{R}\bigg].
\end{eqnarray}

From this relation, Eqs.(\ref{11}) and (\ref{12}) can be re-written as
\begin{eqnarray}\label{a}
3(H^2+\frac{k}{a^2})&=&8\pi G_{eff}(\rho+\rho_{d}),~~~~~~ \\\label{b}
-2(\dot{H}-\frac{k}{a^2})&=&8\pi G_{eff}(\rho+\rho_{d}+p_{d}),
\end{eqnarray}
where $\rho_{d}$ and $p_d$ represent the energy density and the pressure of dark components which are
defined as follows
\begin{eqnarray}\label{c}
\rho_{d}&=&\frac{1}{8\pi G\hat{Y}}[\frac{1}{2}(R\mathcal{F}_{R}-(\mathcal{F}-2\mathcal{F}_{m}L_{m}))
-3H(\dot{R}\mathcal{F}_{RR}+\dot{T}\mathcal{F}_{RT}+\dot{L}_m\mathcal{F}_{RL})],\\\label{d}
p_{d}&=&\frac{1}{8\pi G\hat{Y}}[\frac{-1}{2}(Rf_{R}-(\mathcal{F}-2\mathcal{F}_{m}L_{m}))+2H(\dot{R}\mathcal{F}_{RR}+\dot{T}
\mathcal{F}_{RT}+\dot{L}_m\mathcal{F}_{RL})+\ddot{R}\mathcal{F}_{RR}\nonumber\\
&+&\dot{R}^{2}\mathcal{F}_{RRR}+2\dot{R}\dot{T}\mathcal{F}_{RRT}+\ddot{T}\mathcal{F}_{RT}
+\dot{T}^2 \mathcal{F}_{RTT}+2\dot{R}\dot{L}_m\mathcal{F}_{RRL}+\ddot{L}_{m}\mathcal{F}_{RL}+\dot{L}_m^{2}\mathcal{F}_{RLL}].
\end{eqnarray}
Using Eqs.(\ref{c}) and (\ref{d}), the EoS of dark fluid can be achieved from the relation
$(p_{d}=\omega_{d}\rho_{d})$ and is given by
\begin{eqnarray}\nonumber
\omega_{d}&=&-1+\frac{1}{\frac{1}{2}(R\mathcal{F}_{R}-(\mathcal{F}-2\mathcal{F}_mL_m))-3H(\dot{R}\mathcal{F}_{RR}
+\dot{T}\mathcal{F}_{RT}+\dot{L}_m\mathcal{F}_{RL})}[\ddot{R}\mathcal{F}_{RR}+\dot{R}^{2}\mathcal{F}_{RRR}
+2\dot{R}\dot{T}\mathcal{F}_{RRT}+\\\label{EoS}
&&\ddot{T}\mathcal{F}_{RT}+\dot{T}^{2}\mathcal{F}_{RTT}+2\dot{R}\dot{L}_m\mathcal{F}_{RRL}
+\ddot{L}_{m}\mathcal{F}_{RL}+\dot{L}_m^{2}\mathcal{F}_{RLL}-H(\dot{R}\mathcal{F}_{RR}+\dot{T}\mathcal{F}_{RT}
+\dot{L_m}\mathcal{F}_{RL})].
\end{eqnarray}
Ordinary matter's semi-conservation equation can be written as
\begin{equation}\label{16}
\dot{\rho}+3H(\rho+p)=q.
\end{equation}
The EMT of dark components may also obey the conservation laws which are given as
\begin{equation}\label{17}
\dot{\rho_{d}}+3H(\rho_{d}+p_{d})=q_{d},~~~~~
\end{equation}
\begin{equation}\label{18}
\dot{\rho_{tot}}+3H(\rho_{tot}+p_{tot})=q_{tot},
\end{equation}
where $\rho_{tot}=\rho+\rho_{d}$ , $p_{tot}=p_{d}$ and $q_{tot}=q+q_{d}$.
Here $q_{tot}$ is the term used to express the total amount of energy exchanged in FRW universe
and $q_{d}$ is the energy exchange term of dark components.
It is worthy to mention here that the energy exchange term's relationship in $\mathcal{F}(R,T)$ gravity can be recovered if $L_m=0$ which implies that $\hat{Y}=1+\frac{\mathcal{F}_T}{8\pi G}$. Similarly, by assuming $T=0$ yields $\hat{Y}=1+\frac{\mathcal{F}_L}{16\pi G}$ and results in the total energy term's relationship of $\mathcal{F}(R,L_m)$ gravity. Also, in the limit of GR (when $\mathcal{F}(R,L_{m},T)= R$), the term $q_{tot}$ vanishes.

Substituting Eqs.(\ref{a}) and (\ref{b}) in above equation, we obtain
\begin{eqnarray}\label{19}
q_{tot}&=&\frac{3}{8\pi G}(H^2+\frac{k}{a^2})\partial_{t}(\frac{\mathcal{F}_{R}}{\hat{Y}}).
\end{eqnarray}

\section{Thermodynamical Laws in $\mathcal{F}(R,L_m,T)$ Gravity}

The discovery of thermal behaviour of blackhole established a fundamental link between
thermodynamics and gravity \cite{17a}-\cite{21}. Jacobson \cite{21} demonstrated that by
using the Clausius relationship $TdS=\delta{Q}$ and the entropy proportionality to the horizon area,
it is likely to deduce the Einstein field equations in Rindler spacetime. Here $\delta{Q}$ is the
energy flux all over the horizon and $T$ is the Unruh temperature slightly inside the horizon, as
seen by an accelerated observer. Kofman and Frolov \cite{22} used this approach to compute the field
equations of a steadily rotating foreground scalar field in an inflationary cosmos utilizing quasi de-Sitter
geometry and the relation $-dE= TdS$.

For entropy of the observable horizon by using the first law of thermodynamics, Kim and Cai \cite{23}
constructed the field equations of FRW cosmos with some geometric curvature. After this, Akbar and Cai \cite{32} rewrote the FRW field equations of GR in the pattern given by $dE=TdS+WdV$ at the apparent horizon, where $E=\rho V$ is the overall energy contained within the apparent horizon and $W=\frac{1}{2}(\rho-p)$
represents the work density. In the context of modified theories of gravity like $\mathcal{F}(R)$ theory \cite{R9}, Gauss-Bonnet gravity \cite{32}, Lovelock gravity \cite{34,35}, braneworld gravity \cite{36}, torsion based gravity \cite{37}-\cite{42}, scalar-tensor gravity \cite{34,43,44} and Horava-Lifshitz theory \cite{45*}, the link between gravity and thermodynamic has been investigated by numerous authors and interesting results have been obtained. It is argued that a non-equilibrium thermodynamical description is necessary in $\mathcal{F}(R)$, $\mathcal{F}(R,T)$, $\mathcal{F}(R,L_m)$ as well as scalar-tensor theories of gravity \cite{10m}, \cite{37}-\cite{R11},
where the clausius relationship is replaced by $TdS = \delta{Q}+d\acute{S}$ with $d\acute{S}$ as
the extra entropy production term. Likewise, Bahamonde et al. \cite{45n} explored the non-equilibrium
picture of thermodynamics in the context of $f(T,B)$ theory and presented the corresponding constraints
for validity of second law of thermodynamics.

In this section, we shall formulate first and generalized laws of thermodynamics in the gravitational
framework of $\mathcal{F}(R,L_m,T)$ theory. We shall also explore the validity of corresponding
constraints for some simple $\mathcal{F}(R,L_m,T)$ models.

\subsection{First Law of Thermodynamics}

In this segment, we shall explore the validity of first law of thermodynamics in $\mathcal{F}(R,L_{m},T)$
gravity at the apparent horizon (AH) for non-flat FRW cosmos. The radius of AH is defined by
\begin{equation}\label{20}
\hat{r}_{A}=\frac{1}{\sqrt{H^2+\frac{k}{a^2}}}.
\end{equation}
The temperature related with the AH is defined by using surface gravity $\xi_{sg}$ as
\begin{equation}\label{21}
T_{Ah}= \frac{|\xi_{sg}|}{2\pi},
\end{equation}
where $\xi_{sg}$ is given by
\begin{equation}\label{22}
\xi_{sg}=\frac{-1}{\hat{r}_{A}}(1-\frac{\dot{\hat{r}}_A}{2H\hat{r}_{A}})=-\
\frac{\hat{r}_{A}}{2}(2H^2+\dot{H}+\frac{k}{a^2}).
\end{equation}
In GR, the Bekenstein-Hawking relation calculates the horizon entropy given by $\varepsilon_{h}=\frac{A}{4G}$,
where $A=4\pi\hat{r}^2$ is the area of the AH \cite{17a}-\cite{19}. In modified theory of gravity, the entropy
of black hole solutions with Bifurcate killing horizons is a Noether charge entropy claimed by Wald \cite{W}.
It relies on the variation of Lagrangian density of modified theories with respect to the Reimanian tensor.
The Wald entropy is equivalent to one by four parts of horizon area in units of effective gravitational coupling,
that is,  $\varepsilon_{h}^\prime=\frac{A}{4G_{eff}}$. In $\mathcal{F}(R,L_{m},T)$ gravity, the Wald entropy can
be written as
\begin{equation}\label{23}
\varepsilon_{h}^\prime=\frac{A\mathcal{F}_{R}}{4G\hat{Y}}.
\end{equation}
The above entropy relationship can be reduced to entropy relationship
term's of $\mathcal{F}(R,T)$ or $\mathcal{F}(R,L_m)$ gravities by taking $L_m$ or $T=0$, respectively.
Taking the time derivative of Eq.(\ref{20}) and by using Eq.(\ref{b}), we obtain
\begin{equation}\label{24}
\mathcal{F} _{R}d\hat{r_{A}}=4\pi G\hat{r}^3_{A}(\rho_{tot}+p_{tot})H\hat{Y} dt,
\end{equation}
where $d\hat{r_{A}}$ is the microscopic change in the radius of AH over a time interval $dt$.
Solving Eqs.(\ref{23}) and (\ref{24}), we get
\begin{equation}\label{25}
\frac{1}{2\pi\hat{r_{A}}}d\varepsilon_{h}^\prime=4\pi\hat{r}^3_{A}(\rho_{tot}+p_{tot})Hdt
+\frac{\hat{r}_{A}}{2G\hat{Y}}d\mathcal{F}_{R}+\frac{\hat{r}\mathcal{F}_{R}}{2G}d(\frac{1}{\hat{Y}}).
\end{equation}
Multiplication on both sides of above equation with the factor $(1-\frac{\dot{\dot{\hat{r}}}_A}{2H\hat{r}_{A}})$
leads to the following relationship:
\begin{equation}\label{26}
T_{Ah}d\varepsilon_{h}^\prime=4\pi\hat{r}^3_{A}(\rho_{tot}+p_{tot})Hdt-2\pi\hat{r}^2_{A}(\rho_{tot}
+p_{tot})d\hat{r}_{A}+\frac{\pi\hat{r}^2_{A}T_{Ah}d\mathcal{F}_{R}}{G\hat{Y}}
+\frac{\pi\hat{r}^2_{A}T_{Ah}\mathcal{F}_{R}}{G}d(\frac{1}{\hat{Y}}).
\end{equation}
Now we shall describe the universe's energy within the AH. The Misner-sharp energy is specified
as $E=\frac{\hat{r}_A}{2G}$ and in the gravitational framework of $\mathcal{F}(R,L_{m},T)$ theory,
it takes the following form:
\begin{equation}\label{27}
\hat{E}=\frac{\hat{r}_{A}}{2G_{eff}}.
\end{equation}
In terms of volume $V=\frac{4}{3\pi}\hat{r}^3_{A}$, the above equation can also be written as
\begin{equation}\label{28}
\hat{E}=\frac{3V}{8\pi G_{eff}}(H^2+\frac{k}{a^2})=V_{\rho tot}.
\end{equation}
It denotes the actual energy within the sphere with radius $\hat{r}$. It is clear that $\hat{E}$ is
positive if $G_{eff}=\frac{G\hat{Y}}{\mathcal{F}_{R}}$ is positive and hence the effective gravitational
coupling constant in $\mathcal{F}(R,L_{m},T)$ gravity must be greater than zero, i.e., $G_{eff}>0$.
From Eqs.(\ref{a}) and (\ref{28}), it can be easily written as
\begin{equation}\label{29}
d\hat{E}=-4\pi\hat{r}^3_{A}(\rho_{tot}+p_{tot})H dt+4\pi\hat{r}^2_{A}\rho_{tot}d\hat{r}_{A}
+\frac{\hat{r}_{A}d\mathcal{F}_{R}}{2G\hat{Y}}+\frac{\hat{r}_{A}\mathcal{F}_{R}}{2G}d(\frac{1}{\hat{Y}}).
\end{equation}
Using Eq.(\ref{29}) in (\ref{26}), it yields the following equation
\begin{equation}\label{30}
T_{Ah}d\varepsilon_{h}^\prime=-d\hat{E}^\prime+\hat{W}dV+\frac{(1+2\pi\hat{r}_{A}T_{Ah})\hat{r}_{A}d\mathcal{F}_{R}}{2G\hat{Y}}
+\frac{(1+2\pi\hat{r}_{A}T_{Ah})\hat{r}_{A}\mathcal{F}_{R}}{2G}d(\frac{1}{\hat{Y}})
\end{equation}
in which $\hat{W}=\frac{1}{2}(\rho_{tot}-p_{tot})$. Consequently, the above equation can be re-written as
\begin{equation}\label{31}
T_{Ah}d\varepsilon_{h}^\prime+T_{Ah}d_{i}\varepsilon_{h}^\prime=-d\hat{E}+\hat{W}dV,
\end{equation}
where
\begin{eqnarray}\label{32}
d_{i}\varepsilon_{h}^\prime&=&-\frac{\hat{r}_{A}}{2GT_{Ah}}(1+2\pi\hat{r}_{A}T_{Ah})d(\frac{\mathcal{F}_{R}}{\hat{Y}})
=-\frac{\hat{Y}(\hat{E}+\varepsilon_{h}^\prime T_{Ah})}{T_{Ah}\mathcal{F}_{R}}d(\frac{\mathcal{F}_{R}}{\hat{Y}}).
\end{eqnarray}
It is worthy to mention here that when one equates the cosmological scenario of $\mathcal{F}(R,L_{m},T)$
gravity with GR, Gauss bonnet gravity and Lovelock gravity \cite{32}-\cite{35}, an extra term in first law
of thermodynamics is obtained. This extra term $d_{i}\varepsilon_{h}^\prime$ may be regarded as entropy
production term produced due to the non-equilibrium background of $\mathcal{F}(R,L_{m},T)$ gravity.
If we take $\mathcal{F}(R,L_{m},T)=R$, then the traditional first
law of thermodynamic in GR can be attained. Similarly first law of thermodynamics in $\mathcal{F}(R,T)$ and
$\mathcal{F}(R,L_m)$ can be obtained by taking either $L_m=0$ or $T=0$.

\subsection{Generalized Second Law of Thermodynamics}

During the recent few decades, the validity of GSLT has been investigated in the background of various
modified gravity theories \cite{40}-\cite{43} and interesting results have been achieved. In the present work,
we shall look into its credibility in the newly proposed $\mathcal{F}(R,L_{m},T)$ gravity. According
to GSLT, the time rate of total entropy must be non-negative \cite{43} and hence, one must demonstrate this
by the following mathematical relation:
\begin{equation}\label{33}
\dot{\varepsilon}_{h}^\prime+d_{i}\dot{\varepsilon}_{h}^\prime+\dot{\varepsilon}_{tot}^\prime\geqslant0,
\end{equation}
where $\varepsilon_{h}^\prime$ is the horizon entropy in $\mathcal{F}(R,L_{m},T)$ gravity while
$d_{i}\dot{\varepsilon}_{h}^\prime=\partial_{t}(d_{i}\varepsilon_{h}^\prime)$ is the entropy
containing all energy and matter sources in the interior of horizon. The Gibb's equation that
comprises of all energy and matter fluid is given by \cite{43}
\begin{equation}\label{34}
T_{tot}d\varepsilon_{tot}^\prime=d(\rho_{tot}V)+p_{tot}dV,
\end{equation}
where $T_{tot}$ is the temperature of total energy within the horizon. We consider that $T_{tot}$
is proportional to the temperature of AH, i.e., $T_{tot}=bT_{Ah}$, where we need to impose $0<b<1$
to assure that temperature being non-negative and smaller than the horizon temperature \cite{40}-\cite{43}.
Inserting Eqs.(\ref{31}) and (\ref{34}) in (\ref{33}), we get
\begin{equation}\label{35}
\dot{\varepsilon}_{h}^\prime+d_{i}\dot{\varepsilon}_{h}^\prime+\dot{\varepsilon}_{tot}^\prime
=\frac{24\pi\Theta}{\hat{r}_{A}bR}\geqslant0,
\end{equation}
where
$\Theta=(1-b)\dot{\rho}_{tot}V+(1-\frac{b}{2})(\rho_{tot}+p_{tot})\dot{V}$ is the generalized condition
to preserve the GSLT in modified gravity theories. Using Eqs.(\ref{a}) and (\ref{b}) in the condition (\ref{33}),
it can be written as
\begin{equation}\label{36}
\frac{12\pi\Psi}{bRG\hat{Y}(H^2+\frac{k}{a^2})^2}\geq0,
\end{equation}
where
$$\Psi=2(1-b)H(\dot{H}-\frac{k}{a^2})(H^2+\frac{k}{a^2})\mathcal{F}_{R}+(2-b)H(\dot{H}
-\frac{k}{a^2})^{2}\mathcal{F}_{R}+(1-b)(H^2+\frac{k}{a^2})^{2}\hat{Y}\partial_{t}(\frac{\mathcal{F}_{R}}{\hat{Y}}).$$
As a result, the requirement to establish the validity of GSLT is similar to $\Psi\geq0$. In non-flat Friedmann's cosmos,
the GSLT is valid with the conditions that $\partial_{t}(\frac{\mathcal{F}_{R}}{F})\geq0$, $\dot{H}>0$ and $H>0$.
Also, $\hat{Y}$ and $\mathcal{F}_{R}$ are non-negative for maintaining $E>0$. The condition $b=1$ refers the
situation where the temperature difference among exterior and within the horizon is not changed and
consequently, the GSLT is acceptable only if the following constraint holds:
\begin{equation}\label{37}
\gimel=(\dot{H}-\frac{k}{a^2})^2(\frac{\mathcal{F}_{R}}{\hat{Y}})\geq0.
\end{equation}

Now the condition given by Eq.(\ref{36}) depends on Hubble parameter, its time rate and some derivatives
of the function $\mathcal{F}(R,L_m,T)$. Therefore, we require a Hubble parameter and a $\mathcal{F}(R,L_m,T)$
function in order to check the validity of GSLT. In the upcoming subsections, we shall consider two forms of
$\mathcal{F}(R,L_m,T)$ function with some specific choices of $L_m$. Also, we consider the power law scale factor
for Hubble parameter for simplicity reasons. It is worthy to mention here that power law model helps to
explain full cosmic evolution history including accelerating and decelerating eras.

\subsubsection{{Case-1. The multiplicative case: $\mathcal{F}(R,L_{m},T)=R+\gamma T L_{m}+\sigma, ~~with ~~L_{m}= -\rho$}}

Here we shall discuss the validity of GSLT constraint by taking a simple multiplicative choice
for $\mathcal{F}(R,L_{m},T)$ model along with $L_{m}= -\rho$. It is
worthy to point out here that in comparison to other models available in literature, this model is more
general in nature as it is not possible to recover the $\mathcal{F}(R,L m)$ and $\mathcal{F}(R,T)$ theories
however, it can be reduced to GR case. In order to get some analytical form of
$\rho$, we use the power law form of scale factor, i.e., $a(t)=t^{\alpha}$ (where $\alpha>0$) in the equation
of continuity (\ref{10*}) and consequently, energy density is found as $\rho=\rho_0 t^{-3\alpha(\omega+1)}$.
In order to satisfy full continuity equation, one needs to impose the constraint given by Eq.(\ref{10}) which
leads to a specific value of EoS parameter (for ordinary matter) given by $\omega=-1/3$. Also, Eq.(\ref{37}) can be written
as follows
\begin{equation}
\hat{\gimel}=(\dot{H}-\frac{k}{a^2})^{2}\frac{16\pi G}{16\pi G+3\gamma(\omega-1)\rho}.
\end{equation}
In the present case, the GSLT condition takes the following form
\begin{eqnarray}\label{37*}
GSLT&=&96\pi^{2}H(\dot{H}-\frac{k}{a^{2}})^{2}\frac{\mathcal{F}_{R}}{R(8\pi G+f_{T}+\frac{1}{2}\mathcal{F}_{L})(H^{2}+\frac{k}{a^{2}})^{2}}.
\end{eqnarray}
The graphical representation for the validity of this constraint is presented in the left plot of Figure \textbf{1}.
In this graph, one can see the validity region of GSLT which shows that it is valid for all universes, i.e.,
flat universe ($k=0$), open universe ($k=-1$) and closed universe ($k=1$) for all values of $\gamma<0$. For $\gamma>0$,
it initially violates but during late time, it becomes valid.
\begin{figure}
\centering
\includegraphics[width=0.34\textwidth]{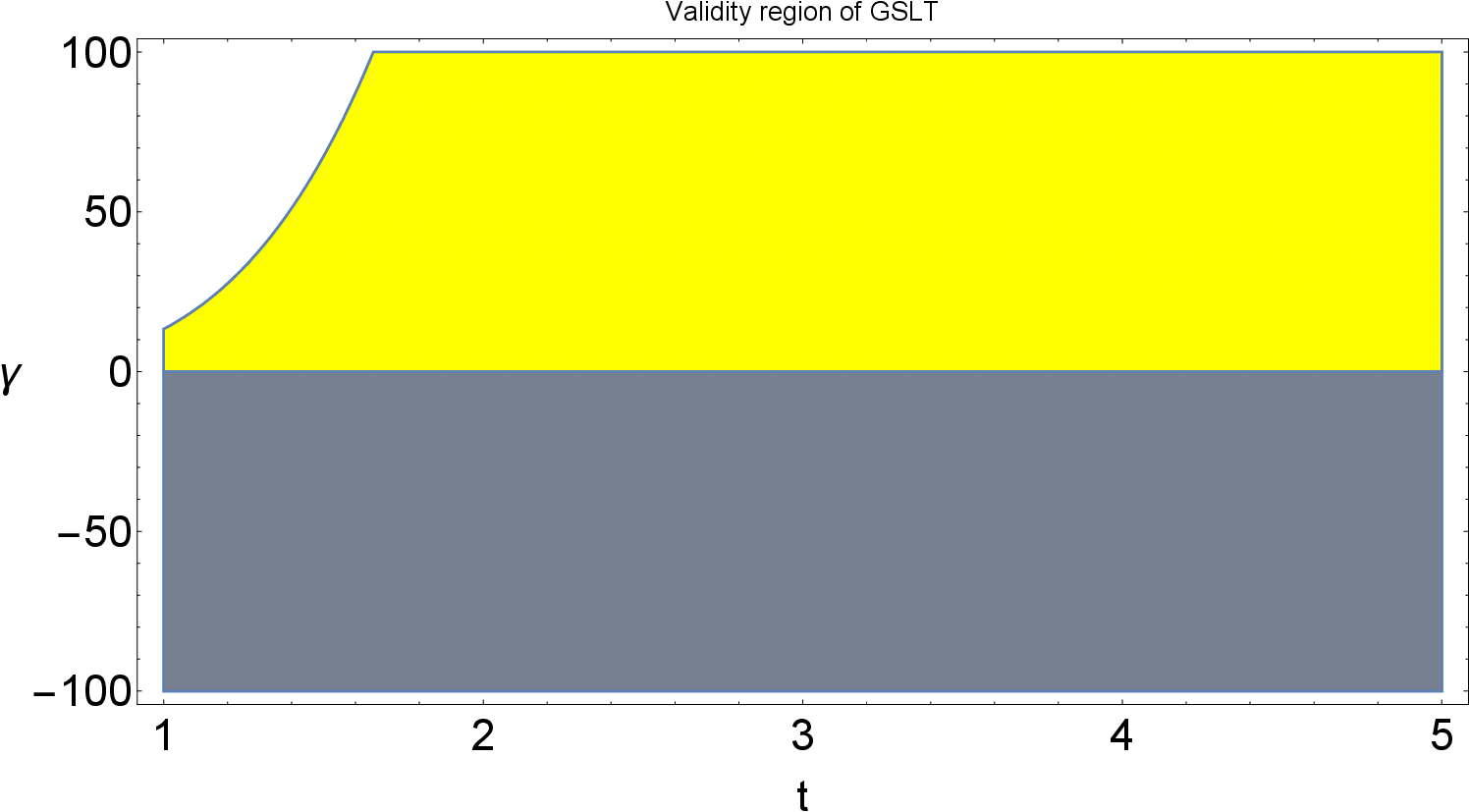}
\includegraphics[width=0.34\textwidth]{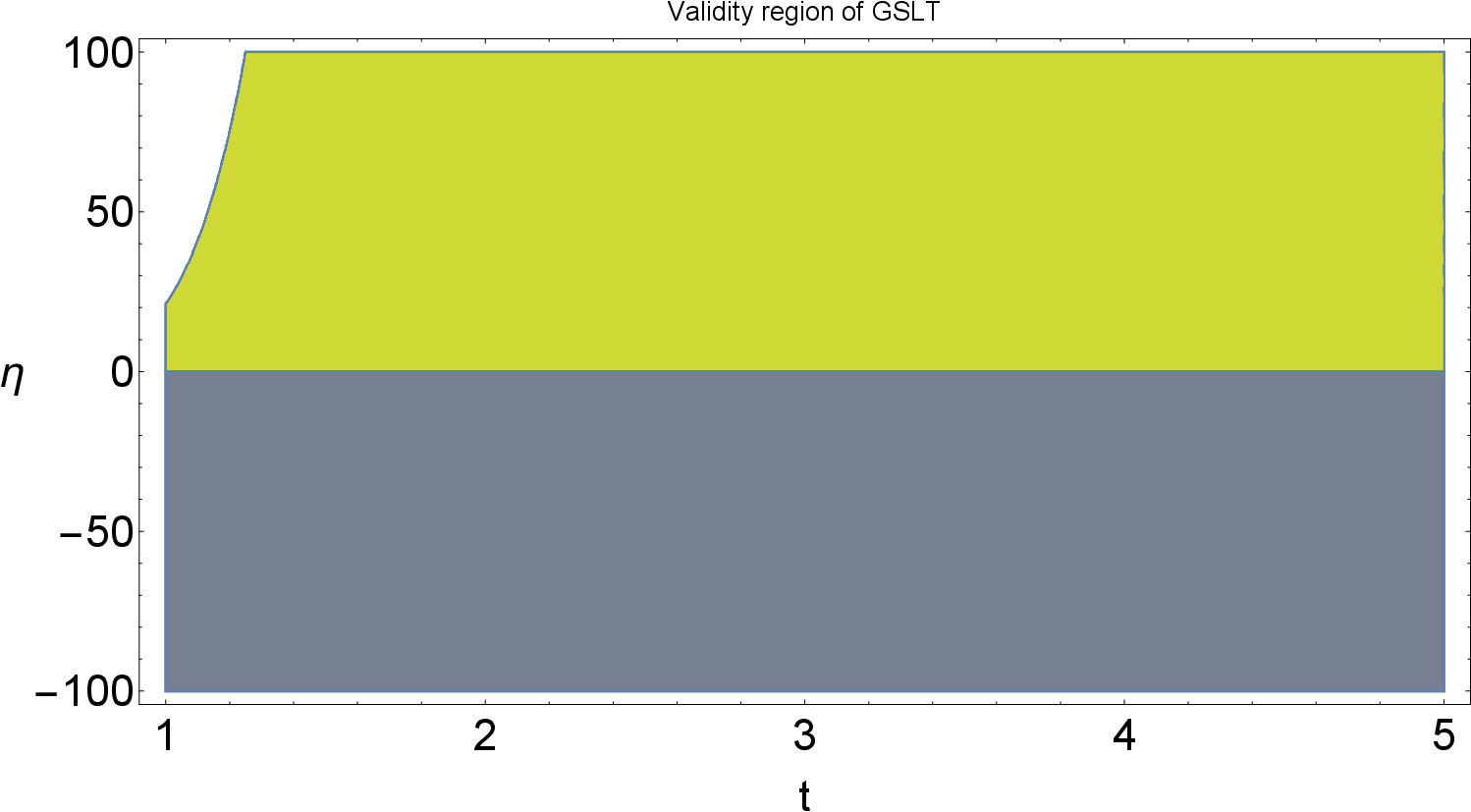}
\caption{\scriptsize{Evolution of GSLT for multiplicative cases with $L_m=-\rho$ and $L_m=p$, respectively. Left plot
indicates graph of GSLT for multiplicative function (case-1) with $L_m=-\rho$ while the right plot represents graph
for multiplicative function (case-2) with $L_m=p$. Here we have fixed free parameters as $\alpha=2$, $\omega=-\frac{1}{3}$
(case-1) and $\omega=.16227$ (case-2).}}\label{fig0}
\end{figure}

\subsubsection{{Case-2. The multiplicative Case: $\mathcal{F}(R,L_{m},T)=R+\eta TL_{m},~~with~~L_{m}= p$}}

Here we shall investigate the validity of GSLT constraint for another interesting choice
$\mathcal{F}(R,L_{m},T)=R+\eta TL_{m}$, where $L_{m}= p$. This model is also
new and it is not possible to recover $\mathcal{F}(R,T)$ and $\mathcal{F}(R,L_m)$ gravities even by
setting $L_m=T=0$. Here the expression for ordinary matter pressure is given by $p=\omega\rho_0t^{-3\alpha(\omega+1)}$,
where the power law scale factor is implemented. In order
to be consistent with the full continuity equation, we use the constraint (\ref{10}) and consequently, it
yields the following equation for $\omega$:
\begin{eqnarray}\nonumber
&&\omega^2+6\omega-1=0.
\end{eqnarray}
The above equation provides two possibilities of EoS parameter: either $\omega=.16227$ or $\omega=-6.16227$.
Further, Eq.(\ref{37}) can be written as
\begin{equation}
\acute{\gimel}=(\dot{H}-\frac{k}{a^2})^{2}\frac{16\pi G}{16\pi G+\eta(5p-\rho)}.
\end{equation}
In this case, the GSLT condition is given by
\begin{eqnarray}\label{37**}
GSLT&=&96\pi^{2}H(\dot{H}-\frac{k}{a^{2}})^{2}\frac{\mathcal{F}_{R}}{R(8\pi G+\mathcal{F}_{T}+\frac{1}{2}\mathcal{F}_{L})(H^{2}+\frac{k}{a^{2}})^{2}}.
\end{eqnarray}
The validity region of GSLT for this case is shown in the right plot of Figure\textbf{1}. It can be visualized
that the trajectories of GSLT remain positive with $\omega=0.16227$, $k=0,~\pm 1$ and for all values of $\eta<0$.
For $\eta\geq 22$, GSLT becomes positive during later times. But for $\omega=-6.16227$, it become negative
and GSLT violates, therefore we use only $\omega=0.16227$ in this case.

\subsubsection{{Case-3. The additive case: $\mathcal{F}(R,L_{m},T)=R+\psi T+2\zeta L_{m},~~with~~L_{m}=-\rho$}}

Here we shall consider simple additive choice of $\mathcal{F}(R,L_{m},T)$ function along $L_{m}=-\rho$ for
investigating the validity of GSLT. In this model, if we set $L_m=T=0$
then it reduces to GR. This model claims $\mathcal{F}(R,T)$ gravity in the limit: $L_m=0$ while
$\mathcal{F}(R,L_m)$ gravity can be achieved by setting $T=0$. Following a similar pattern, an additional constraint
is obtained by applying function $\mathcal{F}(R,L_{m},T)$ in Eq.(\ref{10}) and is given by
\begin{eqnarray}
&&\frac{\delta}{2}(\dot{\rho}(-1+3\omega)+2\dot{\rho})=0
\end{eqnarray}
which implies that $\omega=-\frac{1}{3}$ and further, Eq.(\ref{37}) can be written as
\begin{equation}
\grave{\gimel}=(\dot{H}-\frac{k}{a^2})^{2}\frac{8\pi G}{8\pi G+\psi+\zeta}.
\end{equation}
The corresponding GSLT condition takes the following form:
\begin{eqnarray}\label{37***}
GSLT&=&96\pi^{2}H(\dot{H}-\frac{k}{a^{2}})^{2}\frac{\mathcal{F}_{R}}{R(8\pi G+\mathcal{F}_{T}
+\frac{1}{2}\mathcal{F}_{L})(H^{2}+\frac{k}{a^{2}})^{2}}.
\end{eqnarray}
\begin{figure}
\centering
\includegraphics[width=0.34\textwidth]{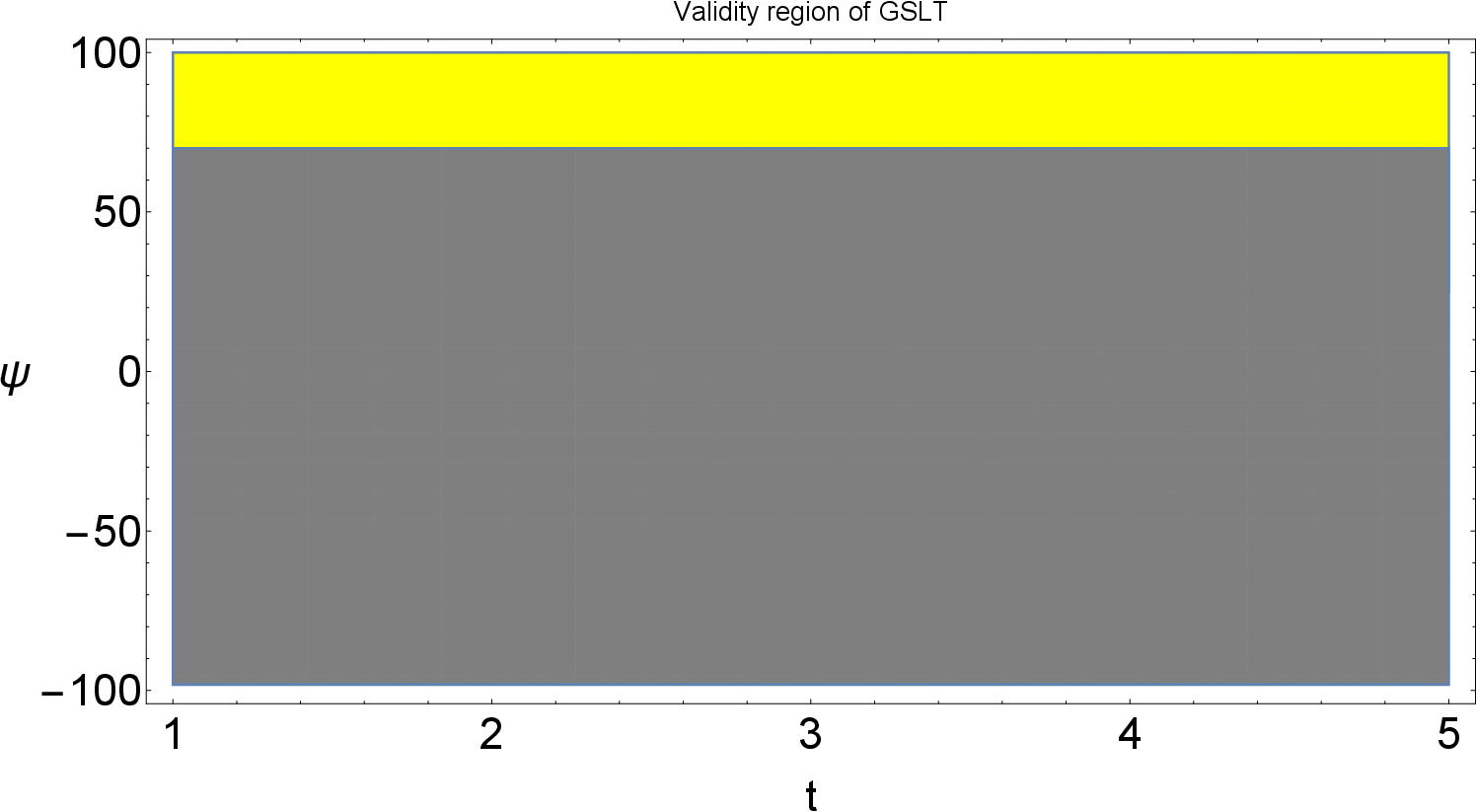}
\caption{\scriptsize{Provides evolution of GSLT for additive case with $L_m=-\rho$. Upper region
shows the validity region for negative $\zeta$ ($\zeta=-50$) and lower region for positive $\zeta$ ($\zeta=73$).
Here we have fixed free parameters as $\alpha=2$ and $\omega=-\frac{1}{3}$.}}\label{fig1}
\end{figure}
In this case, the graphical representation for validity of GSLT constraint is shown in Figure \textbf{2}. This
graph represents the validity region of GSLT for all values of $k$ with $\omega=-1/3$. It is shown that GSLT is
valid in this case for $\{\zeta\geq0$ and $\psi\geq0\}$ or $\{\zeta>72$ and $-100\leq\psi\leq100$\} and
$\{\zeta<0$ and $\psi>72\}$ for flat, open and close universes. It is also seen that validity region for
all universes remains same.

\section{Energy Condition Bounds}

In this subsection, we shall explore the validity of energy condition bounds and behavior of DE
EoS parameter (\ref{EoS}) for the considered models. In the following, we shall define these measures
briefly. For determining the physical significance of any proposed model, the energy condition bounds play
a leading role. These conditions are defined as a set of four constraints (inequalities) namely the weak energy
condition (WEC), the null energy condition (NEC), the strong energy condition (SEC) and the dominant energy
condition (DEC). The formulation of these constraints is purely geometric and these were initially defined
in the framework of Einstein's gravity and later generalized to other gravitational frameworks
(modified gravity theories). In Einstein's gravity, these constraints are defined as
\begin{eqnarray}
NEC:~~~\rho+p\geq0, \quad WEC:~~~\rho\geq0, \quad DEC:~~~\rho-p\geq0, \quad SEC:~~~ \rho+3p\geq0.
\end{eqnarray}
For modified gravitational frameworks, the ordinary density and pressure are replaced by the total
density and pressure (including both ordinary matter and dark energy ingredients). In the present case,
these inequalities are defined by the following relations:
\begin{eqnarray}\nonumber
&& \rho_{tot}+3p_{tot}\geq0 \Rightarrow (\rho+3p)+\frac{1}{8\pi G+\mathcal{F}_{m}}
\left(-(R\mathcal{F}_R-\mathcal{F}+2\mathcal{F}_{m}L_m)+3H(\mathcal{F}_{RR}\dot{R}
+\mathcal{F}_{RT}\dot{T}+\mathcal{F}_{RL}\dot{L}_m)+3\ddot{R}\mathcal{F}_{RR}\right.\\\label{1n}&&\left.+3\dot{R}^2\mathcal{F}_{RRR}
+6\dot{R}\dot{T}\mathcal{F}_{RRT}+3\ddot{T}\mathcal{F}_{RT}+3\dot{T}^2\mathcal{F}_{RTT}
+6\dot{R}\dot{L}_m\mathcal{F}_{RRL}+3\ddot{L}_m\mathcal{F}_{RL}+3{\dot{L}_m}^2\mathcal{F}_{RLL}\right)\geq0,\\\label{n2}
&&\rho_{tot}\geq0 \Rightarrow \rho+\frac{1}{8\pi G+\mathcal{F}_m}\left(\frac{1}{2}(R\mathcal{F}_R-\mathcal{F}
+2\mathcal{F}_mL_m)-3H(\mathcal{F}_{RR}\dot{R}+\mathcal{F}_{RT}\dot{T}+\mathcal{F}_{RL}\dot{L}_m)\right)\geq0,\\\nonumber
&&\rho_{tot}+p_{tot}\geq0 \Rightarrow \frac{1}{8\pi G+\mathcal{F}_m}\left(-H(\mathcal{F}_{RR}\dot{R}
+\mathcal{F}_{RT}\dot{T}+\mathcal{F}_{RL}\dot{L})+\ddot{R}\mathcal{F}_{RR}
+\dot{R}^2\mathcal{F}_{RRR}+2\dot{R}\dot{T}\mathcal{F}_{RTT}+\ddot{T}\mathcal{F}_{RT}\right.\\\label{n3}&&\left.
+2\dot{R}\dot{L}_m\mathcal{F}_{RLL}+\ddot{L}_m\mathcal{F}_{RL}+{\dot{L}_m}^2\mathcal{F}_{RLL}\right)\geq0,\\\nonumber
&&\rho_{tot}-p_{tot}\geq0 \Rightarrow \frac{1}{8\pi G+\mathcal{F}_m}\left((R\mathcal{F}_R-\mathcal{F}+2\mathcal{F}_mL_m)
-5H(\mathcal{F}_{RR}\dot{R}+\mathcal{F}_{RT}\dot{T}+\mathcal{F}_{RL}\dot{L})-\ddot{R}\mathcal{F}_{RR}
-\dot{R}^2\mathcal{F}_{RRR}\right.\\\label{n4}&&
\left.-2\dot{R}\dot{T}\mathcal{F}_{RRT}-\ddot{T}\mathcal{F}_{RT}-\dot{T}^2\mathcal{F}_{RTT}
-2\dot{R}\dot{L}_m\mathcal{F}_{RRL}-\ddot{L}_m\mathcal{F}_{RL}-{\dot{L}_m}^2\mathcal{F}_{RLL}\right)\geq0.
\end{eqnarray}
Now we shall explore the validity of energy conditions for the previously discussed three choices of $\mathcal{F}(R,T,L_m)$
function. For the first case when $\mathcal{F}(R,L_{m},T)=R+\gamma T L_{m}+\sigma, ~~with ~~L_{m}= -\rho$, the above
constraints, respectively, take the following form:
\begin{equation}\label{n5}
\frac{\sigma-2\gamma\rho^2}{8\pi G-\gamma\rho+\frac{\gamma T}{2}}\geq0, \quad \frac{2\gamma\rho^2-\sigma}{16\pi G}\geq0,
\quad \frac{-\sigma+2\rho^2\gamma}{8\pi G-\gamma\rho+\frac{\gamma T}{2}}\geq0.
\end{equation}
It is worthy to mention here that for ordinary matter density and pressure, these condition are satisfied
so one needs to focus only on dark energy terms. Also, it is checked that the NEC constraint $\rho_{tot}+p_{tot}\geq0$
is trivially satisfied for this model. The graphical representation of WEC constraint is provided by the left
plot of Figure \textbf{3} where the validity regions are given. It can be easily checked that WEC holds for
two possible choices of free parameters $-100<\gamma<100,~~\sigma<0$ or $0<\gamma<100,~~\sigma>0$. In the
first case, WEC is valid for $-100<\gamma<100,~~\sigma<0$ for later time while for initial time, WEC can be
valid for $0<\gamma<100,~~\sigma<0$. In the second possibility $0<\gamma<100,~~\sigma>0$, WEC holds for initial
time values only.

\begin{figure}
\centering
\includegraphics[width=0.30\textwidth]{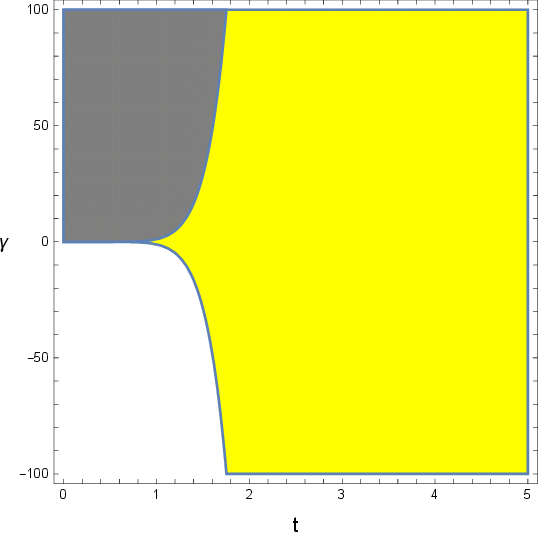}
\includegraphics[width=0.30\textwidth]{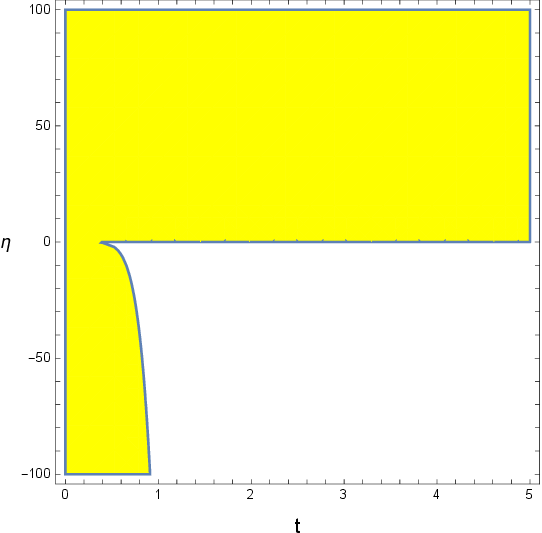}
\includegraphics[width=0.30\textwidth]{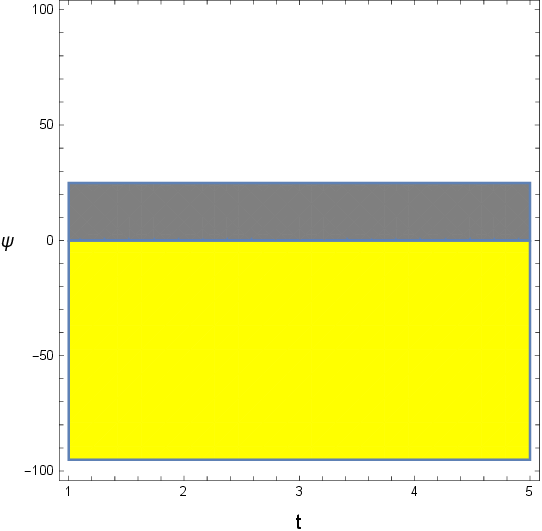}
\caption{\scriptsize{Indicates the validity regions of WEC for different $\mathcal{F}(R,T,L_m)$ cases
with $L_m=-\rho$ and $L_m=p$. Left plot indicates graph of WEC for multiplicative function (case-1) with $L_m=-\rho$,
the middle plot represents graph for multiplicative function (case-2) while the right plot corresponds to additive
function with $L_m=\rho$. Herein, for the left plot, we have fixed free parameters as $\alpha=2,~\omega=-\frac{1}{3},~\rho_0=0.315$,
$\delta=-2$ (for yellow region) and $\delta=2$ (for gray region). For middle plot, we have taken: $\omega=0.16227~\rho_0=0.315$
and $\alpha=2$. While for the right plot, we have chosen $\alpha=2,~\omega=-1/3$, $\zeta=70$ (for yellow region)
and $\zeta=-50$ (for gray region).}}\label{fig3}
\end{figure}
For the second choice $\mathcal{F}(R,L_{m},T)=R+\eta TL_{m},~~with~~L_{m}= p$, these energy constraints
can be written as
\begin{equation}\label{n6}
\frac{-2\eta\omega^2\rho^2}{8\pi G+\eta\omega\rho+\frac{\eta T}{2}}\geq0, \quad
\frac{2\eta\omega^2\rho^2}{16\pi G+\eta\rho(1-\omega)}\geq0, \quad
\frac{2\eta\omega^2\rho^2}{8\pi G+\eta\omega\rho+\frac{\eta T}{2}}.
\end{equation}
While the constraint corresponding to NEC trivially holds in this case. The graphical illustration of WEC
constraints in terms of validity regions is provided in the middle plot of Figure \textbf{3}. It indicates that
the inequality for WEC holds when free parameters satisfy $0<\eta<100$ and $\alpha>0$. For the range $0>\eta>-100$,
the NEC can be valid only for initial times $0<t<1$.

For the case $\mathcal{F}(R,L_{m},T)=R+\psi T+2\zeta L_{m}$ with $L_{m}=-\rho$, these constraints can be
defined as
\begin{equation}\label{n7}
\frac{2\rho\psi+\psi T}{8\pi G+\psi+\zeta}\geq0, \quad \frac{-3\psi\rho(1-\omega)}{16\pi G+2\psi+2\zeta}\geq0,
\quad \frac{(\rho(3\omega-2)-1)}{8\pi G+\psi+\zeta}.
\end{equation}
Likewise, the inequality referring to NEC trivially holds. The graphical representation of constraint for WEC
is given by right plot of Figure \textbf{3}. It can be easily seen that this constraint is valid for either
$0<\psi<30,~ -100<\zeta<-50$ or when $-100<\psi<0,~ 70<\zeta<100$.

Now we shall discuss the behavior of EoS parameter for DE for the discussed model and determine
the possible constraints on the free parameters. It is well-known that the EoS parameter for DE
represents the quintessence era of cosmic evolution when $-1<\omega_d<0$ while it refers to phantom
epoch if $\omega_d<-1$. Further, for positive values, it corresponds to matter dominated era. For the first case,
the DE EoS parameter takes the following form:
\begin{eqnarray}\nonumber
\omega_d=-1+\frac{-\frac{113.4k\gamma}{t^{10}}-\frac{204.12\gamma}{t^8}}{\frac{0.893025\gamma}{t^8}-0.5\sigma},
\end{eqnarray}
where we have fixed other parameters. For flat universe, it is seen that quintessence region can be achieved
by fixing either $\gamma<0,~~~\sigma<0$ or $\gamma>0,~~~\sigma>0$. Also, it is checked that $\omega_d=-1$ if
we set $\gamma=0$. Further, the phantom cosmic epochs can be achieved when $\sigma>0~~~\gamma<0$ or $\sigma<0~~~\gamma>0$.
In the second case, the DE EoS parameter takes the form:
\begin{eqnarray}\nonumber
\omega_d=-1+\frac{73.9508t^{-4+1.48681\alpha}\alpha(kt^2(1+2\alpha)+t^{2\alpha}(-3+6\alpha))}{\rho_0}.
\end{eqnarray}
It is interesting to mention here that in the above expression, the model parameter $\gamma$ disappears.
It can be easily checked that for open, closed or flat universes, the above expression can only yield
quintessence era of cosmic evolution by picking very small values of $\alpha$ or $\rho_0$
(which seems inappropriate) and hence the EoS refers to matter dominated eras. For the third choice of
$\mathcal{F}(R,T,L_m)$ function, the DE EoS parameter can refer to phantom phase of cosmic evolution if one pick
$\psi<0,~~~\alpha>1$ with $k=0$. The corresponding expression is given by
\begin{eqnarray}\nonumber
\omega_d=-1+\frac{2(-12kt^{-2-2\alpha}(-1-2\alpha)\alpha
+\frac{36\alpha(-1+2\alpha)}{t^4})\psi}{2t^{-2\alpha}\psi\rho_0+2t^{-2\alpha}\zeta\rho_0-2t^{-2\alpha}(\psi+\zeta)\rho_0}.
\end{eqnarray}

\section{Reconstruction}

In this section, we shall adopt the well known reconstruction scheme to reconstruct the form of
generic function $\mathcal{F}(R,L_{m},T)$ by taking power law model and $\Lambda$CDM model in flat universe
and de Sitter model in non-flat universe.

\subsection{Power Law Model}

Here we shall reconstruct the $\mathcal{F}(R,L_m,T)$ function by taking power law model ($a(t)=a_0t^{\alpha}$)
into account. The Hubble parameter $H$ and Ricci scalar $R$ in flat universe are then given by
\begin{equation}
H=\frac{\alpha}{t} \longrightarrow R=6\alpha(-1+2\alpha)t^{-2}.
\end{equation}
Here $\alpha >0$. Using the power law of scale factor along with the Hubble parameter in Eq.(\ref{8}),
we obtain
\begin{equation}\label{p1}
8\pi G\rho+\frac{1}{2}(\mathcal{F}_{L}+2\mathcal{F}_{T})(\rho+L_{m})-\frac{\mathcal{F}}{2}
+\frac{\alpha-1}{2(-1+2\alpha)}R\mathcal{F}_{R}+\frac{R^2}{-1+2\alpha}\mathcal{F}_{RR}
+\frac{3\alpha(1+w)}{2(-1+2\alpha)}RT\mathcal{F}_{RT}-3H\dot{L_{m}}\mathcal{F}_{RL}=0.
\end{equation}
Equation (\ref{p1}) is difficult to solve analytically for the unknown function $\mathcal{F}(R,L_{m},T)$,
so we consider a simple choice of the function as separable sum form given by
$\mathcal{F}(R,L_{m},T)=\lambda_{1}\mathcal{F}_1(R)+\lambda_{2}\mathcal{G}(L_{m})+\lambda_{3}\mathcal{H}(T)$,
where we choose $\mathcal{G}(L_m)=L_m=\rho$ to find the solution of involved unknown functions. After
inserting this form in Eq.(\ref{p1}), we obtain a differential equation whose solution can be found by using
constraint (\ref{10}) and is given by
\begin{eqnarray}\nonumber
\mathcal{F}(R,L_{m},T)&=&R^{\frac{1}{4}(3-\alpha-\sqrt{1+10\alpha+\alpha^2})}\lambda_{1}a_{1}
+R^{\frac{1}{4}(3-\alpha+\sqrt{1+10\alpha+\alpha^2})}\lambda_{1}a_{2}+\lambda_{2}L_m\\\label{P8}&&
T^{\frac{1}{2}+\frac{\sqrt{4+3(\omega-1)(3\omega-1)}}{4}}a_3\lambda_3+T^{\frac{1}{2}
-\frac{\sqrt{4+3(\omega-1)(3\omega-1)}}{4}}a_4\lambda_3+\frac{T (16 \pi G+\lambda _2)}{ (3 \omega-1)},
\end{eqnarray}
where $a_{i},~~i=1,2,3,4$ are constants of integration. The above reconstructed
model can be reduced to reconstructed model for GR by setting $L_m=T=0$. Similarly it can claim reconstructed model
for $F(R,L_m)$ and $F(R,T)$ \cite{Sum1} gravities by inserting $T=0$ and $L_m=0$, respectively. Now we check the validity
of GSLT for power law reconstructed solution with different values of coupling parameters $\lambda_1,~\lambda_2$ and $\lambda_3$.
For this, we divide it into three different choices of $\lambda_i,~~i=1,2,3$. Firstly, we choose
$\lambda_1=0$ $\Rightarrow$ $GSLT=0$, so it does not give any information about validity of GSLT,
therefore $\lambda_1\neq 0$. Secondly, we choose $\lambda_2=0$ and find the ranges of $\lambda_1$ and $\lambda_3$.
It is to be found that GSLT is valid for $\{\lambda_1\geq 0$ and $\lambda_3\geq 0\}$ or $\{\lambda_1 \leq 0$ and
$\lambda_3\leq-74.67\}$. Lastly, for $\lambda_3=0$ the GSLT is valid for $\{\lambda_1\leq0$ and
$\lambda_2\leq -50.27\}$ or $\{\lambda_1\geq 0$ and $\lambda_2\geq-50.26\}$. The validity regions for
different values of $\lambda_i$ are shown in Figure \textbf{4}. In both plots, the yellow region represent
the validity region of GSLT for $\lambda_1<0$ and blue region for $\lambda_1>0$.
\begin{figure}
\centering
\includegraphics[width=0.34\textwidth]{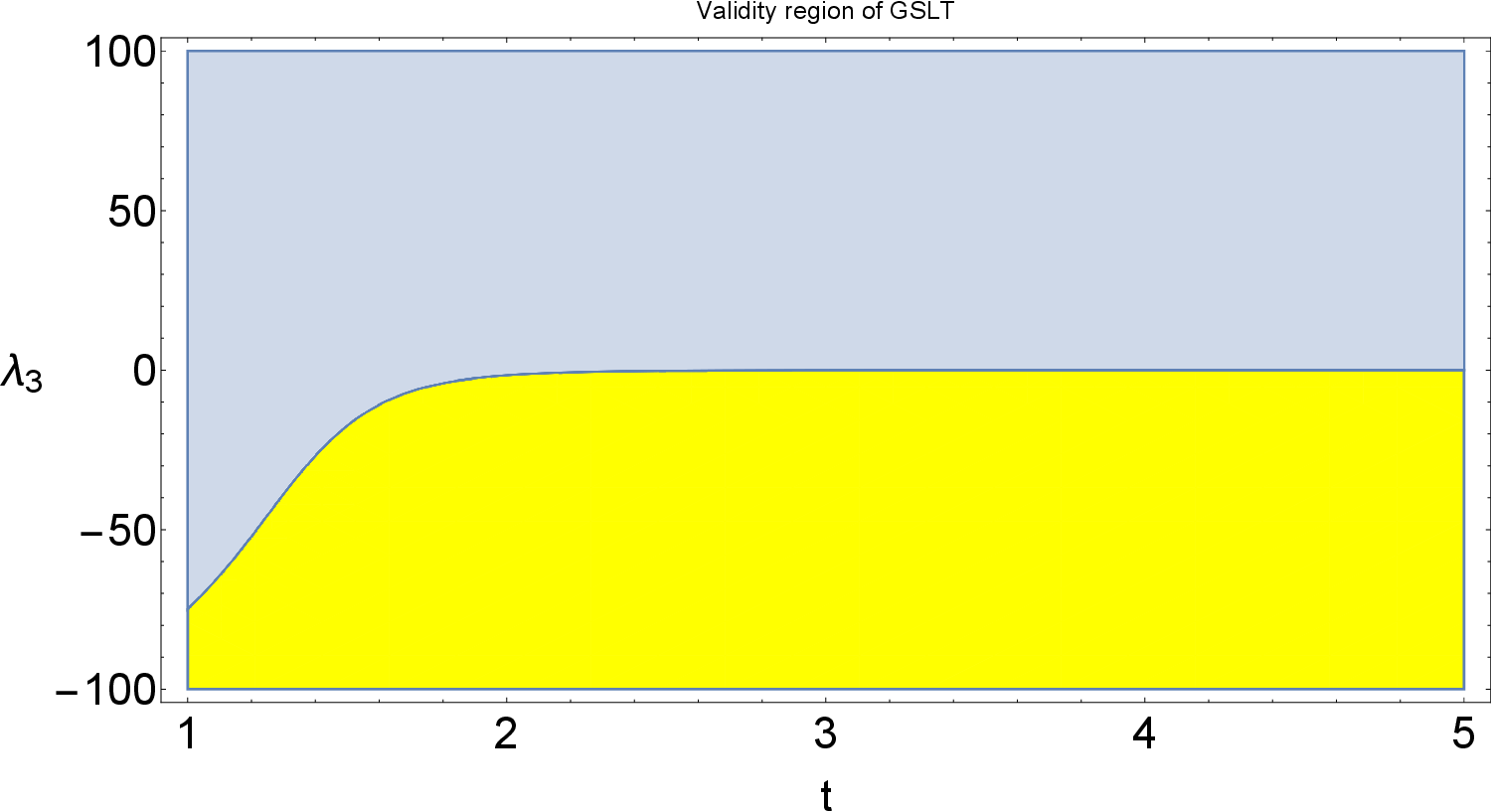}
\includegraphics[width=0.34\textwidth]{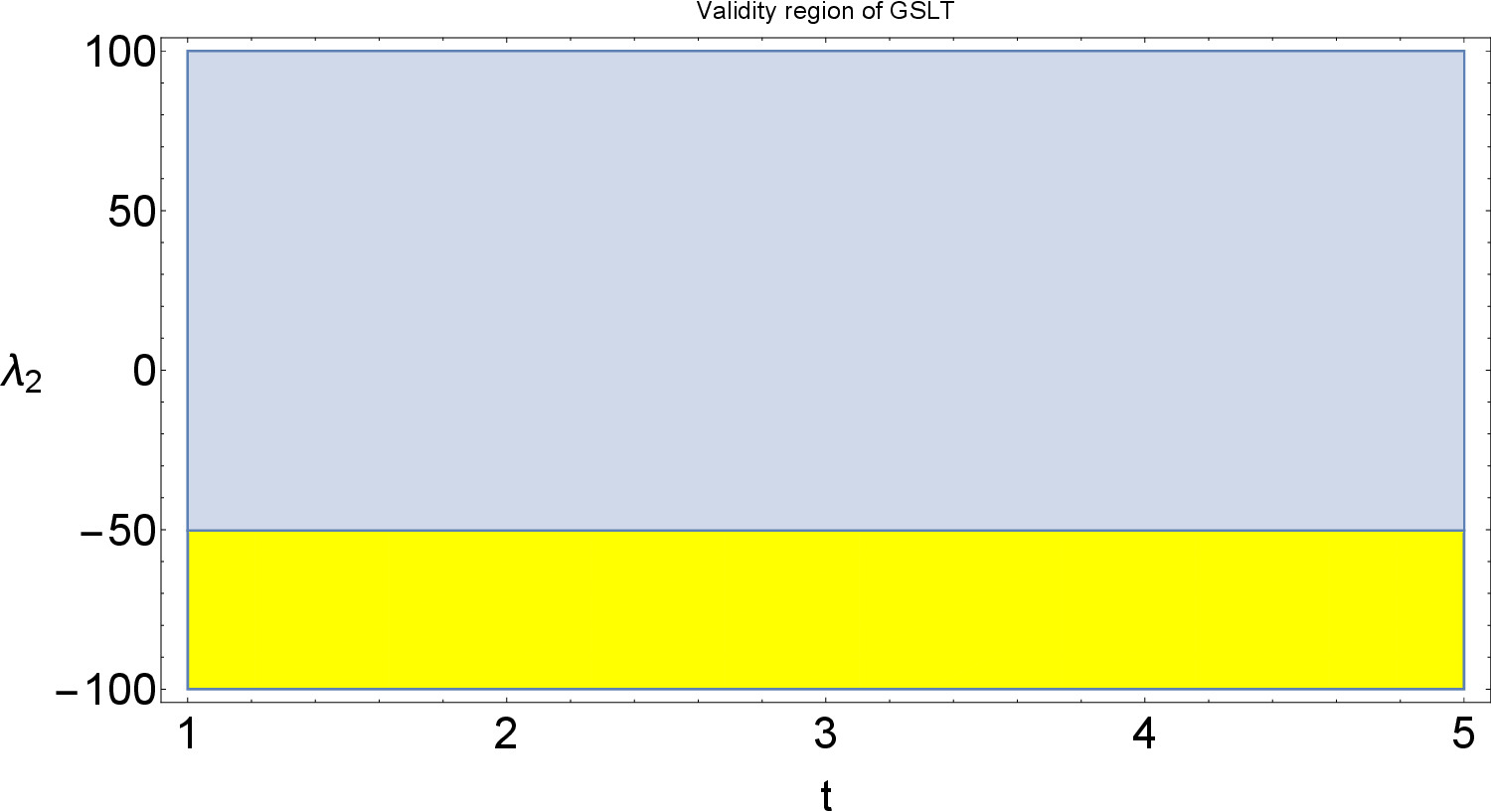}
\caption{\scriptsize{Evolution of GSLT for reconstructed power law model with $L_m=\rho$. The left plot
represents the validity region of GSLT for $\lambda_2=0$ and the right plot indicates the validity region
of GSLT for $\lambda_3=0$. In both plots for $\lambda_1<0$ (yellow and $\lambda_1=-10$) and $\lambda_1>0$
(blue and $\lambda_1=10$) In right figure yellow plot represents the validity region of GSLT when $\lambda_1<0$
and $\lambda_3=-50.27$, while blue plot indicates the region for $\lambda_1\geq 0$ and $\lambda_2=10$.
Here we have fixed free parameters as $\alpha=2,~\omega=\frac{2}{3}$ and $a_1=a_2=a_3=a_4=1$.}}\label{fig2}
\end{figure}
\begin{figure}
\centering
\includegraphics[width=0.34\textwidth]{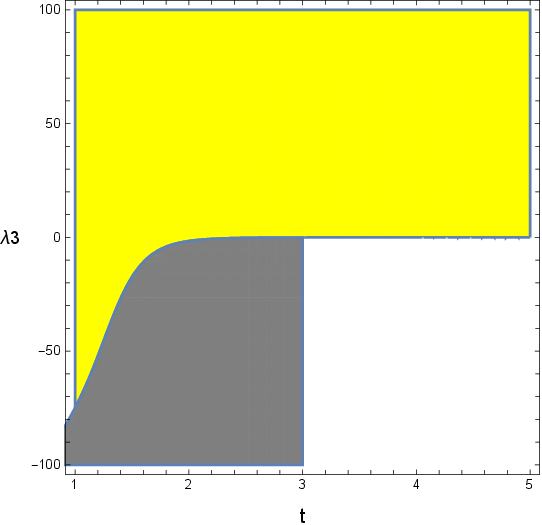}
\caption{\scriptsize{Evolution of energy condition for reconstructed power law model with $L_m=\rho$.
Here we have fixed free parameters as $\alpha=2,~\lambda_2=0,~\omega=\frac{2}{3}$ and $a_1=a_2=a_3=a_4=1$. Also,
for yellow and grey regions, we have fixed $\lambda_1=40$ and $\lambda_1=-40$, respectively.}}\label{fig2}
\end{figure}
In Figure \textbf{5}, the validity region of WEC for power law reconstructed model is provided where two
possibilities of free constant are obtained: $\lambda_2=0,~~40<\lambda_1<100,~~\lambda_3>0$
and $100<\lambda_1<-40,~~\lambda_3<0$. The expression for EoS parameter for DE parameter (\ref{EoS}) takes
quite lengthy expression in terms of $\lambda_i$ and cosmic time. Since we are interested to explore the values of
coupling constants $\lambda_2$ and $\lambda_3$ for which the DE dominated cosmic epochs (quintessence/phantom)
can be obtained so we set $\lambda_1=0$. The 3D plot of the corresponding expression indicates that
quintessence epoch requires $\lambda_3=0$ while $\lambda_2$ can be any real number while phantom era
can be achieved in later times for any non-zero values of $\lambda_2$ and $\lambda_3$. It is also interesting
to mention here that for getting $\omega_d=-1$, one must impose
\begin{eqnarray}
\lambda_3=\frac{\left(\frac{1}{t^{10}}\right)^{0.933}(-3.92141\times10^{19}
-7.8014\times10^{17}\lambda_2)}{5.50919\times10^{16}+7.30644\times10^{17}\left(\frac{1}{t^{10}}\right)^{0.866}}.
\end{eqnarray}

\subsection{$\Lambda$CDM Model}

In this segment, we shall first reconstruct $\mathcal{F}(R,L_{m},T)$ model, which can represent the $\Lambda$CDM
cosmic period while taking dust into account as a matter content. The first FRW equation describes the relevant Hubble
parameter as \cite{45}
\begin{equation}\label{L1}
H=\sqrt{{H_0}^2+\frac{\kappa^2}{3}\rho_0a^{-3}}=\sqrt{{H_0}^2+\frac{\kappa^2}{3}\rho_0a_{0}^{-3}e^{-3N}}.
\end{equation}
The initial portion of Eq.(\ref{L1}) on the right side represents the cosmological constant, while the
remainder reflects the observable and CDM contributions. Using the relation $H=\frac{\dot{a}}{a}=\frac{dN}{dt}$,
the Ricci scalar $R$ is given by $R=6(HH'+2H^2)$ while time rates of matter Lagrangian and EMT tensor trace take
the form: $\dot{T}=HT'$ and $\dot{L_{m}}=HL_{m}'$, respectively. Here prime denotes the derivative with respect to $N$.
Applying these values of Hubble parameter and $R$ in Eq.(\ref{8}), we obtain the following differential equation:
\begin{equation}\label{L2}
8\pi G\rho+\frac{1}{2}(\mathcal{F}_L+2\mathcal{F}_T)(\rho+L_m)-\frac{\mathcal{F}}{2}+3(HH'+H^2)\mathcal{F}_R
-3H[6(H^2H''+HH'^2+4H^2H')\mathcal{F}_{RR})+3HT'\mathcal{F}{RT}+HL_m'\mathcal{F}_{RL}]=0.
\end{equation}
Now we apply $\mathcal{F}(R,L_m,T)=\lambda_1\mathcal{F}_1(R)+\lambda_2L_m+\lambda_3\mathcal{H}(T)$, along with $L_m=\rho$
in above equation and then it can be separated into two differential equations as follows
\begin{eqnarray}\label{L3}
&&18H(H^2H''+HH'^2+4H^2H')\mathcal{F}_{1RR}-3(HH'+H^2)\mathcal{F}_{1R}+\frac{1}{2}\mathcal{F}_1(R)=0,\\\label{L4}
&&\frac{(16\pi G+\lambda_2)}{2(-1+3\omega)}T+\frac{2\lambda_3T}{(-1+3\omega)}\mathcal{H}_T-\frac{\lambda_3}{2}\mathcal{H}(T)=0,
\end{eqnarray}
To solve the first differential equation, we use the Hubble parameter in terms of Ricci scalar given by
\begin{equation}\nonumber
H^2=-(\frac{R}{3}+3H_0^2).
\end{equation}
Using this expression along with the following relations:
\begin{equation}\nonumber
~~~HH'=\frac{3}{2}(H_0^2-H^2), ~~~HH''=-3HH'-H'^2
\end{equation}
in Eq.(\ref{L3}), it can be written as
\begin{equation}
3(R+9H_0^2)(R+12H_0^2)\mathcal{F}_{1RR}-(\frac{R}{2}+9H_0^2)\mathcal{F}_{1R}+\frac{1}{2}\mathcal{F}_1(R)=0,
\end{equation}
To solve above equation, we define a new variable $z=\frac{R}{3H_0^2}+4$ and consequently, this equation
can be changed to the following form
\begin{equation}
z(1-z)\mathcal{F}_{1zz}+[\sigma-(\alpha+\beta+1)z]\mathcal{F}_{1z}-\alpha\beta\mathcal{F}_1(R)=0,
\end{equation}
where $\sigma=-\frac{1}{3}$, $\alpha+\beta+\sigma=-\frac{1}{6}$, $\alpha\beta=-\frac{1}{6}$.
The above mentioned equation is a hypergeometric equation whose solution is provided by
\begin{equation}
\mathcal{F}_1(z)=\beta_{1}F(\alpha,\beta,\sigma;z)+\beta_{2}z^{1-\sigma}F(\alpha-\sigma+1,\beta-\sigma+1,2-\sigma;z).
\end{equation}
Thus $\mathcal{F}(R,L_m,T)$ model can be constituted as
\begin{eqnarray}\nonumber
\mathcal{F}(R,L_m,T)&=&\lambda_1\beta_{1}F(\alpha,\beta,\sigma;\frac{R}{3H_0^2}+4)
+\lambda_1\beta_2(\frac{R}{3H_0^2}+4)^{1-\sigma}F(\alpha-\sigma+1,\beta-\sigma+1,2-\sigma;\frac{R}{3H_0^2}+4)
+\lambda_{2}L_m\\\label{L5}
&&+T^{\frac{1}{2}+\frac{\sqrt{4+3(\omega-1)(3\omega-1)}}{4}}a_3\lambda_3+T^{\frac{1}{2}
-\frac{\sqrt{4+3(\omega-1)(3\omega-1)}}{4}}a_4\lambda_3+\frac{T (16 \pi G+\lambda _2)}{ (3 \omega-1)},
\end{eqnarray}
where $\beta_i's$ are the constants of integration. By setting $L_m=0$,
it reduces to reconstructed model for $\mathcal{F}(R,T)$ gravity \cite{Sum1}. Again, we use same choices of
$\lambda_i$ to visualize the validity region of GSLT for $\Lambda$CDM reconstructed solution. Since $\lambda_1\neq 0$,
it can either take negative or positive values. For $\lambda_2=0$, the GSLT is valid for $\{\lambda_1\leq 0$ and
$-100\leq\lambda_3 \leq100\}$ or $\{\lambda_1\geq0$ and $\lambda_3<0\}$. For $\lambda_3=0$, then GSLT
is valid for $\{\lambda_1\leq 0$ and $\lambda_2\geq -50\}$ or $\{\lambda_1\geq 0$ and $\lambda_2<-51\}$.
The validity regions of GSLT for different values of $\lambda_i$ are represented in Figure \textbf{6}.
\begin{figure}
\centering
\includegraphics[width=0.34\textwidth]{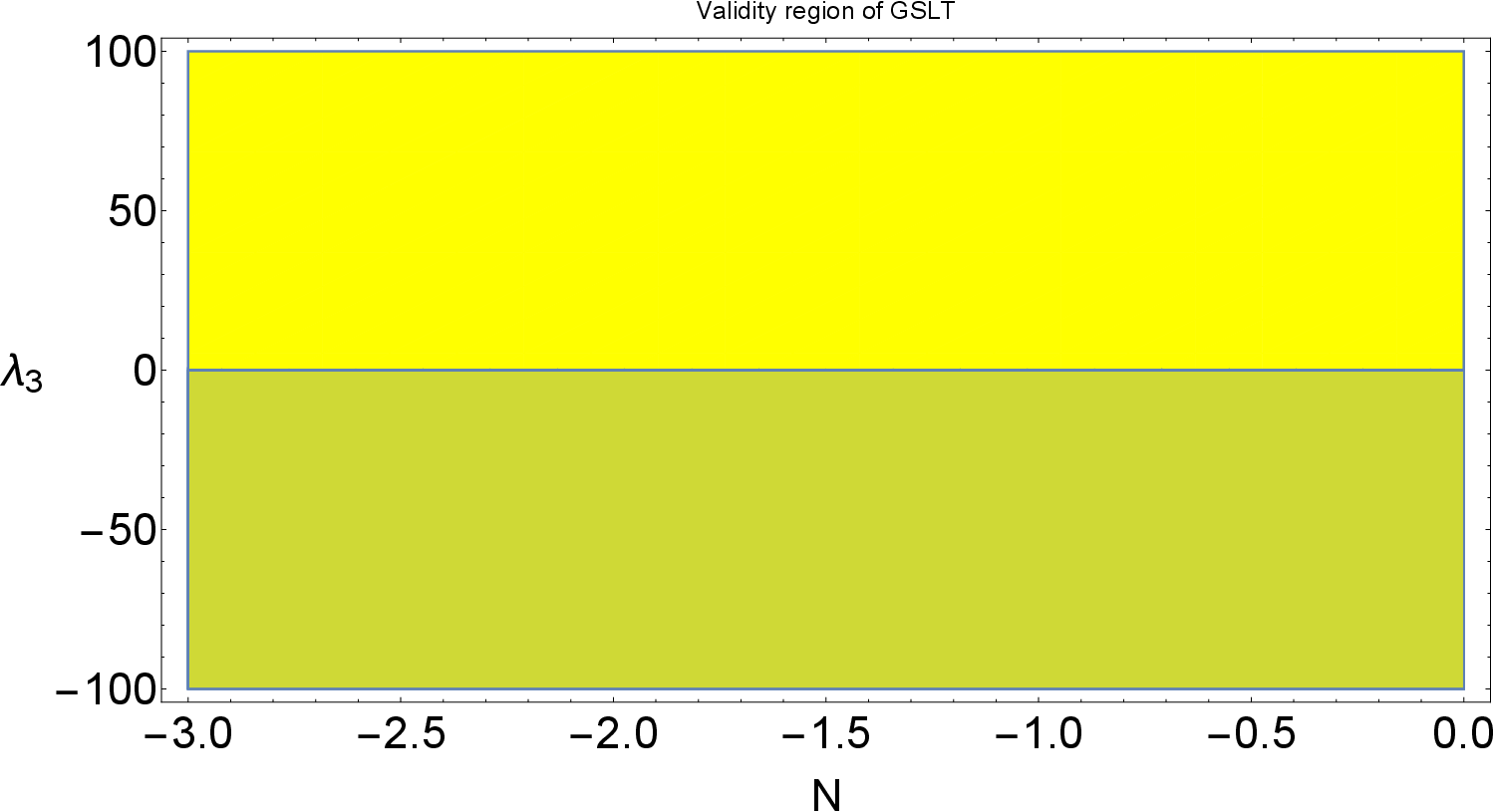}
\includegraphics[width=0.34\textwidth]{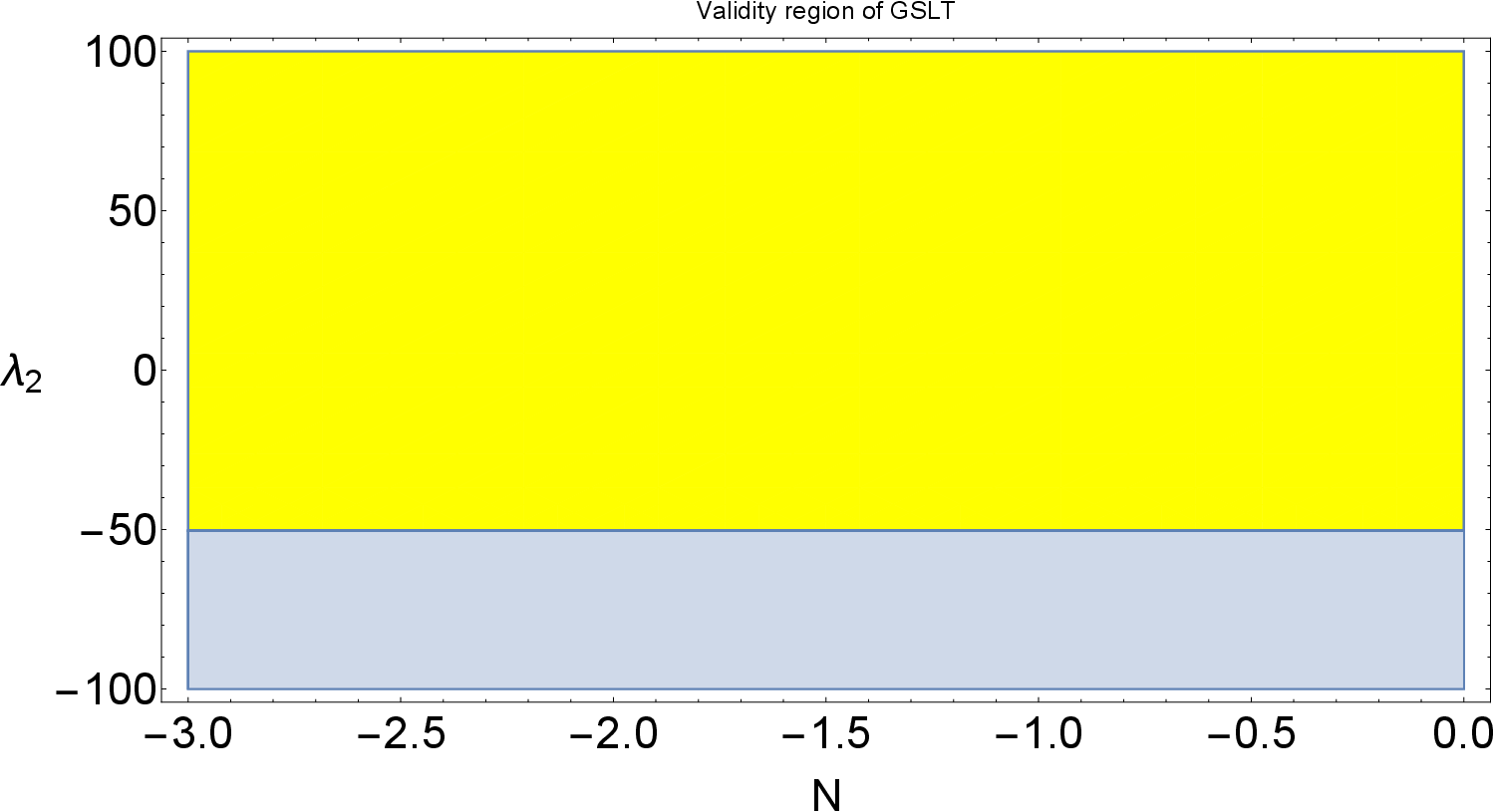}
\caption{\scriptsize{Evolution of GSLT for reconstructed $\Lambda$CDM model with $L_m=\rho$. The left plot
represents the validity region of GSLT for $\lambda_2=0$ and right plot indicates the validity region of GSLT
for $\lambda_3=0$. Here $\lambda_1=-10$ (for $\lambda_1<0$) and $\lambda_1=10$ (for $\lambda_1\geq 0$).
Here we have fixed free parameters as $\omega=\frac{2}{3}$ and $\beta_1=\beta_2=a_3=a_4=1$.}}\label{fig3}
\end{figure}
\begin{figure}
\centering
\includegraphics[width=0.34\textwidth]{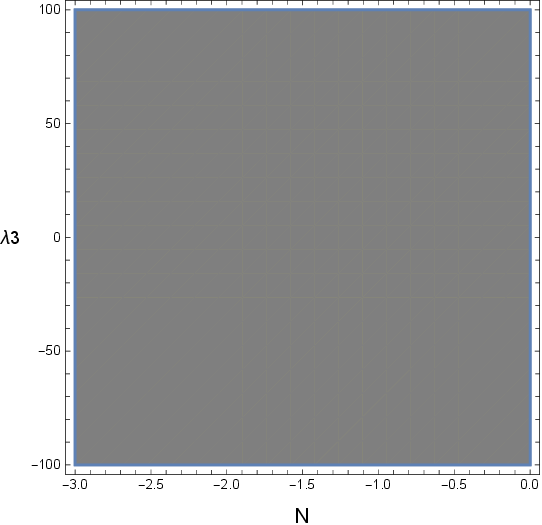}
\caption{\scriptsize{Evolution of WEC for reconstructed $\Lambda$CDM model with $L_m=\rho$ is shown.
Here we have fixed free parameters as $\alpha=2,~\omega=\frac{2}{3},~\lambda_1=0$ and $a_1=a_2=a_3=a_4=1$.}}\label{fig2}
\end{figure}
The validity region for WEC constraint is provided in Figure \textbf{7}. It is seen that $\lambda_2=0$ cannot be
possible so we have set $\lambda_1=0$ and following possibilities of $\lambda_2$ and $\lambda_3$ are achieved: either
$-15<\lambda_2<0,~~-100<\lambda_3<-80,~~ -1<N<0$ or $-30<\lambda_2<-10,~~-100<\lambda_3<-80,~~ -2.5<N<-1.5$.
Further, we compute the corresponding EoS for DE and it is seen that the resulting expression is quite lengthy
involving $\lambda_i$ and e-folding parameter. Since we are interested to explore the impact of $T$ and $L_m$ parts,
so we assume $\lambda_1=0$ and check the 3D region plots for $\omega_d>-1$ or $\omega_d<-1$. It is seen that
quintessence can be attained by considering all values of $\lambda_2$ and $\lambda_3$ ($\lambda_3\neq0$).
While phantom requires $\lambda_3=0$ ($\lambda_2$ can take any real value) in final stages of cosmic evolution.

\subsection{de Sitter universe}

The de Sitter solution plays a significant role in cosmology as it can be used to explain the current era of cosmos.
The scale factor distinguishes the dS model which is defined as $a(t)=a_{0}e^{H_{0}t}$. This is growing exponentially
and results in a constant Hubble parameter $H=H_{0}$ and $R=6(2H_{0}^2+\frac{k}{a_0^2e^{2H_0t}})$ in case of non-flat
universe. In this process, we assume $\mathcal{F}(R,L_m,T)=\lambda_1\mathcal{F}_1(R)+\lambda_2L_m+\lambda_3\mathcal{H}(T)$,
$L_m=\rho$ and the matter source with fixed EoS parameter $\omega=\frac{p}{\rho}$ so that
\begin{equation}\nonumber
\rho=\rho_{0}e^{-3(1+\omega)H_{0}t}, \omega\neq-1.
\end{equation}
Inserting the above relations in Eq.(\ref{8}), we get
\begin{eqnarray}\label{d1}
&&8\pi G\rho+(\lambda_2+2\lambda_3\mathcal{H}_T)\frac{(\rho+L_m)}{2}-\frac{\lambda_1\mathcal{F}_1(R)}{2}-\frac{\lambda_2L_m}{2}
-\frac{\lambda_3\mathcal{H}(T)}{2}+3H_{0}^2\lambda_1\mathcal{F}_{1R}+6H_0^2(R-12H_0^2)\mathcal{F}_{1RR}=0.
\end{eqnarray}
The solution for de Sitter model is given by
\begin{eqnarray}\nonumber
\mathcal{F}(R,L_{m},T)&=&\lambda_1(c_1 \cos [\frac{\sqrt{4 H_0^2-\frac{R}{3}}}{H_0}]-c_2\text{Sin}[\frac{\sqrt{4 H_0^2-\frac{R}{3}}}{H_0}])+\lambda_2L_m+T^{\frac{1}{2}+\frac{\sqrt{4+3(\omega-1)(3\omega-1)}}{4}}a_3\lambda_3
+\\\label{d5}&&T^{\frac{1}{2}-\frac{\sqrt{4+3(\omega-1)(3\omega-1)}}{4}}a_4\lambda_3+\frac{T (16 \pi G+\lambda _2)}{ (3 \omega-1)}.
\end{eqnarray}
where $c_1$ and $c_2$ are the constants of integration. Next we investigate the validity region of
GSLT for different values of $\lambda_i$. For $\lambda_2=0$, the GSLT is valid for $\{\lambda_1\leq 0$ and
$\lambda_3\leq 0\}$ or $\{\lambda_1\geq 0$ and $\lambda_3\geq 0\}$. Similarly for $\lambda_3=0$, it is valid
for $\{\lambda_1\geq 0$ and $-50\leq\lambda_2\}$ or $\{\lambda_1\leq 0$ and $\lambda_2\leq-51\}$. The graphical
representation of GSLT for different values of $\lambda_i$ are shown in Figure \textbf{8}.
\begin{figure}
\centering
\includegraphics[width=0.34\textwidth]{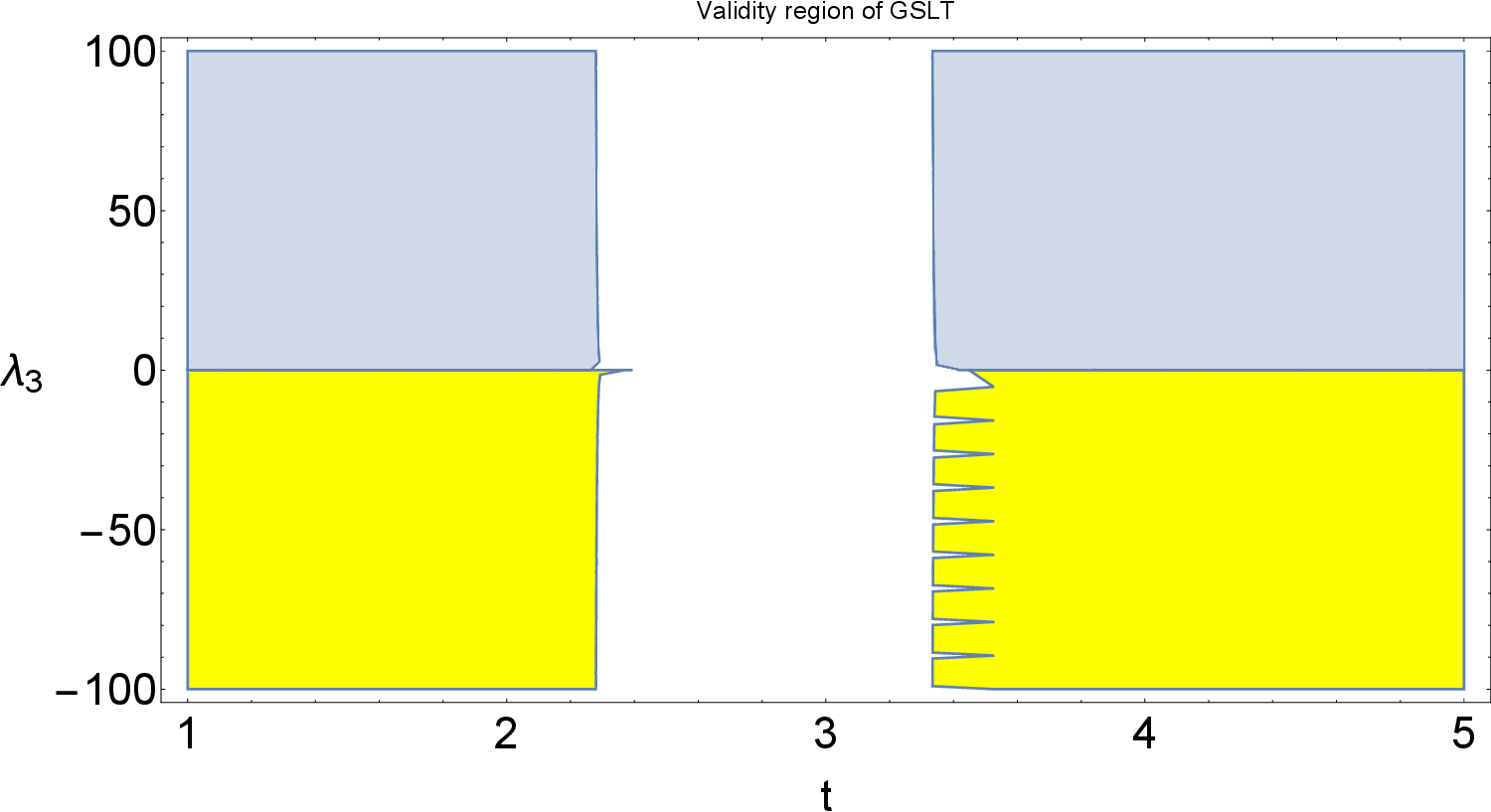}
\includegraphics[width=0.34\textwidth]{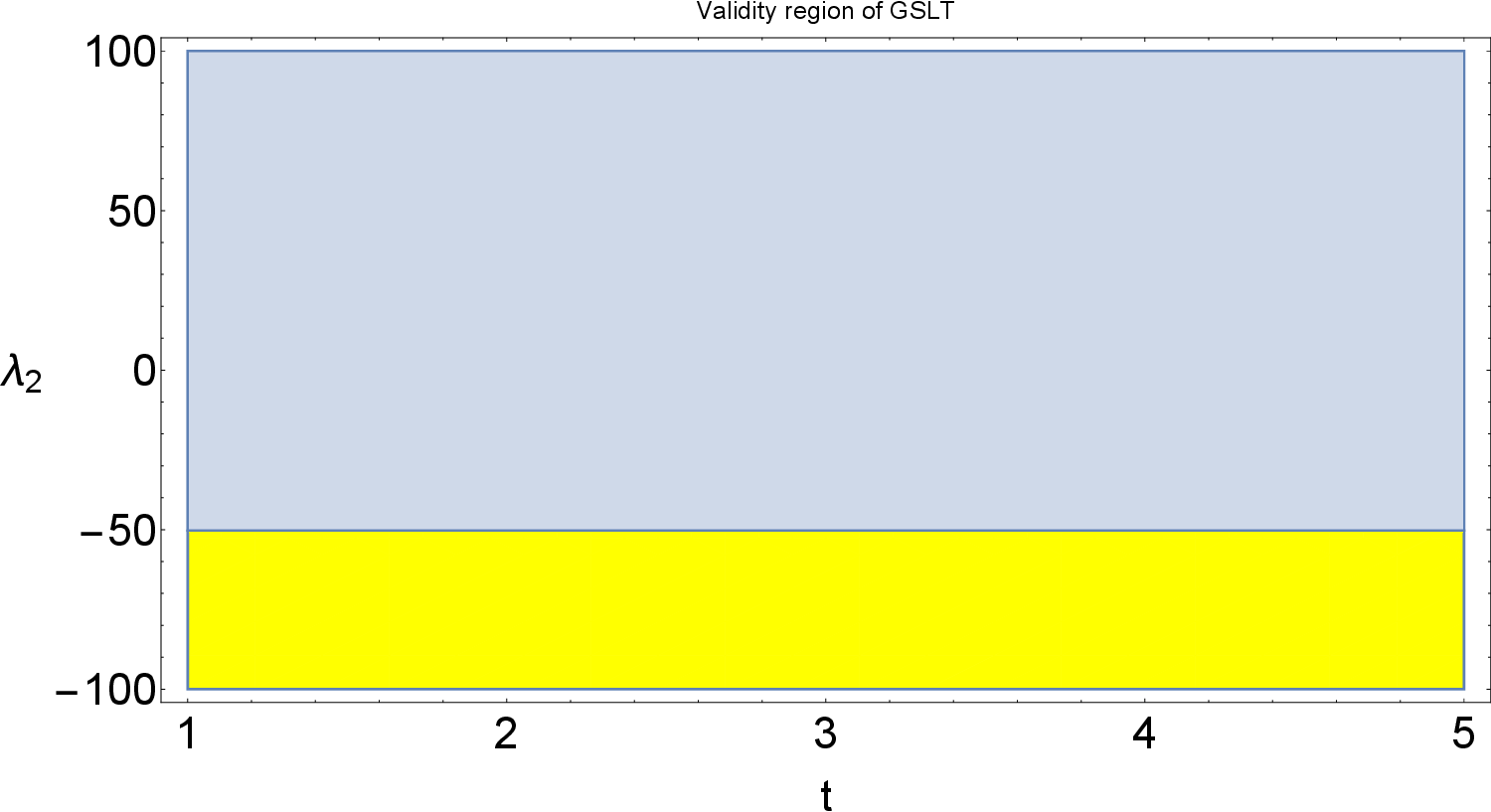}
\caption{\scriptsize{Evolution of GSLT for reconstructed de Sitter universe model with $L_m=\rho$. The left plot
represents the validity region of GSLT for $\lambda_2=0$ and $\lambda_3=-10$ (yellow) or $\lambda_3=10$ (blue).
The right plot indicates the validity region of GSLT for $\lambda_3=0$ and $\lambda_2=-10$ (yellow) and $\lambda_2=10$ (blue).
Here we have fixed free parameters as $\omega=\frac{2}{3},~k=-1$ and $c_1=c_2=a_3=a_4=1$.}}\label{fig4}
\end{figure}
\begin{figure}
\centering
\includegraphics[width=0.34\textwidth]{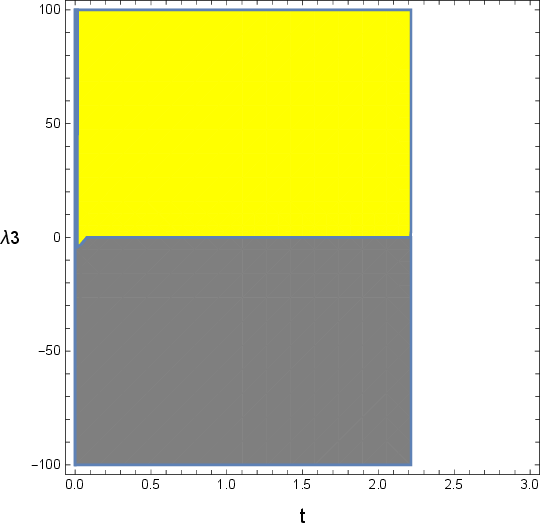}
\caption{\scriptsize{Evolution of energy condition for reconstructed de Sitter model with $L_m=\rho$ is provided.
Here we have fixed free parameters as $\alpha=2,~\omega=\frac{2}{3},~\lambda_2=0$ and $a_1=a_2=a_3=a_4=1$. Also,
we have set $\lambda_1=10$ and $\lambda_1=-10$ for yellow and blue regions, respectively.}}\label{fig2}
\end{figure}
For WEC, the graphical illustration of constraint is provided in Figure \textbf{9}. It is seen that
the inequality holds when $\lambda_2=0,~~0<\lambda_1<50,~~0<\lambda_3<100,$ for $0<t<2$. Likewise,
it is valid when $\lambda_2=0,~~-100<\lambda_1<0,~~-100<\lambda_3<0$ for $0<t<2$. In this case,
it is easy to check that the EoS parameter for DE (\ref{EoS}) leads to value $-1$ on simplification as expected.

\section{Perturbations and Stability}

In this section, we shall analyze the stability of reconstructed power law, $\Lambda$CDM and de Sitter models
in $\mathcal{F}(R,L_m, T)$ gravity. In this regard, we shall find the corresponding perturbed field
equations and continuity equation using FRW universe model for these solutions. Let us consider a solution
\begin{eqnarray}\label{s1}
H(t)=H_{\ast}(t)
\end{eqnarray}
compatible with the background field equations which further leads to form of Ricci scalar given by
$\mathcal{R}_{\star}=6H_{\star}(H_{\ast}^{'}+2H_{\ast})$, where prime indicates derivative with respect to
e-folding parameter. If we consider a particular $F(\mathcal{R},L_m,T)$ model which can reproduce the
solution given in Eq.(\ref{s1}), then Eq.(\ref{5}) takes the following form
\begin{eqnarray}\nonumber
&&8\pi G\rho_{\ast}+(\rho_{\ast}+L_m) \frac{\mathcal{F}_{L}^{\ast}}{2}+(\rho_{\ast}+L_m)\mathcal{F}_{T}^{\ast}
-\frac{\mathcal{F}^{\ast}}{2}+3(H_{\ast}H_{\ast}^{'}+H_{\ast}^{2})\mathcal{F}_{R}^{\ast}
-3H_{\ast}[(6H_{\ast}^2H_{\ast}^{''}+H_{\ast}H_{\ast}^{'^2}+4H_{\ast}^2H_{\ast}^{'})\mathcal{F}_{RR}^{\ast}\\\label{s2}
&&+H_{\ast}T_{\ast}^{'}\mathcal{F}_{RT}^{\ast}+H_{\ast}L_{\ast}^{'}\mathcal{F}_{RL}^{\ast}]=0
\end{eqnarray}
and from Eq.(\ref{9}), the evolution of energy density is given by
\begin{eqnarray}\label{s3}
\rho_{\ast}^{'}+3(1+\omega)\rho_{\ast}&=&\frac{-1}{8\pi +\mathcal{F}_{m}^{\ast}}\bigg[(\rho_{\ast}
+L_m^{\ast})(\mathcal{F}_{m}^{\ast})^{'}-\frac{1}{2}\mathcal{F}_{T}^{\ast}(T_{\ast}^{'}
-2(L_m^{\ast})^{'})\bigg],
\end{eqnarray}
where $\mathcal{F}^{\ast}$ indicates the function corresponding to solution given in Eq.(\ref{s1}).
In case the usual conservation law holds (generally it does not remain valid in $\mathcal{F}(R,L_m,T)$ gravity),
we find $\rho$ in terms of $H_{\ast}(t)$ as follows
\begin{eqnarray}\nonumber
\rho_{\ast}(t)=\rho_{0}e^{(-3(1+\omega)\int H_{\ast}(t)dt)}=\rho_{0}e^{(-3(1+\omega)N)}.
\end{eqnarray}
Further, the perturbed $H(t)$ and $\rho(t)$ are given by
\begin{eqnarray}\label{s4}
H(t)=H_{\ast}(t)(1+\delta(t)), ~~~~~~~~~~~ \rho(t)=\rho_{\ast}(t)(1+\delta_m(t)),
\end{eqnarray}
where $\delta_m(t)$ and $\delta(t)$ represent the perturbation functions. In order to analyze introduced perturbations
about Eq.(\ref{s1}), let us expand $\mathcal{F}(R,L_m,T)$ function as a series in terms of Ricci scalar and
energy-momentum tensor trace and is given by
\begin{eqnarray}\label{s5}
\mathcal{F}(R,L_m,T)=\mathcal{F}^{\ast}+\mathcal{F}_{R}^{\ast}(R-R_{\ast})
+\mathcal{F}_{L}^{\ast}(L_m-L_{m}^{\ast})+\mathcal{F}_{T}^{\ast}(T-T_{\ast})+O^2.
\end{eqnarray}
Here only the linear terms are to be considered for further calculations, while $O^2$ indicating the
quadratic or higher power terms of $R$ and $T$ will be ignored. Using Eqs.(\ref{s4}) and (\ref{s5}) in
FRW equation, we obtain the following perturbed field equation
\begin{eqnarray}\label{s6}
b_2\delta^{''}+b_1\delta^{'}+b_0\delta=c_{m1}\delta_m+c_{m2}\delta_m^{'},
\end{eqnarray}
where $b_i$'s and $c_{mi}$'s are given in Appendix \textbf{I}. Applying above perturbations
in Eq.(\ref{9}), we obtain the following perturbed continuity equation
\begin{eqnarray}\label{s7}
d_1\delta_m^{'}+d_2\delta_m+d_3\delta^{'}+d_4\delta=0,
\end{eqnarray}
where $d_i$'s are also listed in Appendix \textbf{I}. If the usual conservation law holds,
then Eq.(\ref{s6}) reduces to
\begin{eqnarray}\label{s8}
\delta_m^{'}+3\delta=0.
\end{eqnarray}
These perturbed Eqs.(\ref{s6}) and (\ref{s7}) will play the role of key equations in
analyzing the stability of FRW model. For $\mathcal{F}(R,L_m,T)=\lambda_1\mathcal{F}(R)
+\lambda_2\mathcal{G}(L_m)+\lambda_3\mathcal{H}(T)$ type models, these perturbed equations reduce to
\begin{eqnarray}\label{s9}
\hat{b_2}\delta^{''}+\hat{b_1}\delta^{'}+\hat{b_0}\delta=\hat{c_{m1}}\delta_m,~~~~~~~~~~
\hat{d_1}\delta_m^{'}+\hat{d_2}\delta_m+\hat{d_3}\delta=0,
\end{eqnarray}
where $b_i's$ and $d_i's$, ${i=0,1,2,3}$ are listed in Appendix \textbf{I}. In following sections,
we present the stability of power law, de sitter and $\Lambda$CDM models for this form of generic function.

\subsection{Stability of Power Law Solution}

Here, we shall analyze the stability of matter dominated and late time epochs of power law solution.
Now, we consider the reconstructed power law model given in Eq.(\ref{P8}) and perturbed \textbf{ Eq.(\ref{s4})}.
\begin{eqnarray}\nonumber
&&-18H_{\ast}^4\mathcal{F}_{RR}^{\ast}\delta^{''}+\{12\rho_{\ast}H_{\ast}^2\mathcal{F}_{R T}^{\ast}
-(18H_{\ast}^3 H_{\ast}^{'}+54H_{\ast}^4)\mathcal{F}_{RR}^{\ast}-108H_{\ast}^3(H_{\ast}^2H_{\ast}^{''}
+H_{\ast}H_{\ast}^{'^2}+4H_{\ast}^2H_{\ast}^{'})\mathcal{F}_{RRR}^{\ast}-18(-1+3\omega)\\\nonumber
&&H_{\ast}^4\rho_{\ast}^{'}\mathcal{F}_{RRT}^{\ast}+6\rho_{\ast}H_{\ast}^2\mathcal{F}_{R L}^{\ast}
-18H_{\ast}^4\rho_{\ast}^{'}\mathcal{F}_{RRL}^{\ast}\}\delta^{'}+\{-6H_{\ast}^2\mathcal{F}_{R}^{\ast}
+(6\rho_{\ast}(H_{\ast}H_{\ast}^{'}+4H_{\ast}^2)-3H_{\ast}^2\rho_{\ast}^{'})\mathcal{F}_{RL}^{\ast}
-18(H_{\ast}^3H_{\ast}^{'}+4H_{\ast}^4)\\\nonumber &&\rho_{\ast}^{'}\mathcal{F}_{RRL}^{\ast}
+(12\rho_{\ast}(H_{\ast}H_{\ast}^{'}+4H_{\ast}^2)-3(-1+3\omega)H_{\ast}^2\rho_{\ast}^{'})\mathcal{F}_{RT}^{\ast}
-18(2H_{\ast}^3H_{\ast}^{''}+H_{\ast}^2H_{\ast}^{'^2}+7H_{\ast}^3H_{\ast}^{'}-4H_{\ast}^4)\mathcal{F}_{RR}^{\ast}
-108(H_{\ast}^3H_{\ast}^{''}\\\nonumber &&+H_{\ast}^2H_{\ast}^{'^2}+ 4H_{\ast}^3H_{\star}^{'})(H_{\ast}H_{\ast}^{'}
+4H_{\ast}^2)\mathcal{F}_{RRR}^{\ast}-18(H_{\ast}^3H_{\ast}^{'}+4H_{\ast}^4)\rho_{\ast}^{'}(-1+3\omega)\mathcal{F}_{RRT}^{\ast}\}\delta
+\{8\pi G\rho_{\ast}+\frac{3(1+\omega)\rho_{\ast}\mathcal{F}_{T}^{\ast}}{2}+\\\nonumber
&&2\rho_{\ast}^2(-1+3\omega)\mathcal{F}_{TT}^{\ast}-(-1+3\omega)(\rho_{\ast}(3H_{\ast}H_{\ast}^{'}
+H_{\ast}^{2})+3H_{\ast}^2\rho_{\ast}^{'})\mathcal{F}_{RT}^{\ast}-18(3\omega-1)\rho_{\ast}(H_{\ast}^3H_{\ast}^{''}
+H_{\ast}^2H_{\ast}^{'^2}+4H_{\ast}^3H_{\ast}^{'})\mathcal{F}_{RRT}^{\ast}\\\nonumber
&&-3H_{\ast}^2(-1+3\omega)^2\rho_{\ast}\rho_{\ast}^{'}\mathcal{F}_{RTT}^{\ast}
+(1+3\omega)\rho_{\ast}^2\mathcal{F}_{LT}^{\ast}+\frac{1}{2}\rho_{\ast}\mathcal{F}_{L}^{\ast}
+3((H_{\ast}^{'}H_{\ast}+H_{\ast}^2)\rho_{\ast}-H_{\ast}^2\rho_{\ast}^{'})\mathcal{F}_{RL}^{\ast}
+\rho_{\ast}^2\mathcal{F}_{LL}^{\ast}-3H_{\ast}\rho_{\ast}\mathcal{F}_{LRR}^{\ast}\\\label{s10}
&&-6(-1+3\omega)H_{\ast}^2\rho_{\ast}\rho_{\ast}^{'}\mathcal{F}_{LRT}^{\ast}
-3H^2\rho_{\ast}\rho_{\ast}^{'}\mathcal{F}_{L{R}L}^{\ast}\}\delta_{m}
-3H_{\ast}^2((-1+3\omega)\mathcal{F}_{{R}T}^{\ast}+\mathcal{F}_{{R}L}^{\ast})\delta_{m}^{'}=0,
\end{eqnarray}
To solve this equation and Eq.(\ref{s8}), we use numerical technique and yields the following equation:
\begin{eqnarray}\nonumber
&&-18H_{\ast}^4\mathcal{F}_{RR}^{\ast}\delta^{''}+\{-(18H_{\ast}^3 H_{\ast}^{'}+54H_{\ast}^4)\mathcal{F}_{RR}^{\ast}
-108H_{\ast}^3(H_{\ast}^2 H_{\ast}^{''}+H_{\ast}H_{\ast}^{'^2}+4H_{\ast}^2H_{\ast}^{'})\mathcal{F}_{RRR}^{\ast}\}
\delta^{'}+\{-6H_{\ast}^2\mathcal{F}_{R}^{\ast}\\\nonumber
&&-18(2H_{\ast}^3H_{\ast}^{''}+H_{\ast}^2H_{\ast}^{'^2}+7H_{\ast}^3H_{\ast}^{'}
-4H_{\ast}^4)\mathcal{F}_{RR}^{\ast}-108(H_{\ast}^3H_{\ast}^{''}+H_{\ast}^2H_{\ast}^{'^2}
+4H_{\ast}^3H_{\ast}^{'})(H_{\ast}H_{\ast}^{'}+4H_{\ast}^2)\mathcal{F}_{RRR}^{\ast}\}\delta\\\label{s11}
&&+\{8\pi G\rho_{\ast}+\frac{3(1+\omega)\rho_{\ast}\mathcal{F}_{T}^{\ast}}{2}+2\rho_{\ast}^2(-1+3\omega)\mathcal{F}_{TT}^{\ast}
+\frac{1}{2}\rho_{\ast}\mathcal{F}_{L}^{\ast}+\rho_{\ast}^2\mathcal{F}_{LL}^{\ast}\}\delta_{m}=0.
\end{eqnarray}
Here, we are assuming the energy momentum tensor to be conserved covariantly and $\mathcal{F}(R,L_m,T)$ function
given in Eq.(\ref{P8}). We set $\lambda_1=10,~\lambda_2=2$ and $\lambda_3=5$ and solve the Eqs.(\ref{s8}) and
(\ref{s11}) numerically. Initially, we deal with dominated era for $\alpha=2/3$ and analyze the stability of
power law model by plotting graph of $\delta$ and $\delta_m$ for different initial conditions. In (\textbf{a}),
we set $\delta^{'}=\delta_{m}=.1$ and vary $\delta$, in (\textbf{b}), fix $\delta=\delta_{m}=.1$ and
vary $\delta^{'}$and in (\textbf{c}) set $\delta=\delta^{'}=.1$ and vary $\delta_m$ as shown in Figure \textbf{10}.
It is shown that model is stable as it converges to zero. Similarly, we plot the perturbed parameters
$(\delta(N),\delta_m(N))$ by fixing initial conditions in the range $\{-.1,.1\}$ for accelerating era
$\alpha=2$ and analyze the stability of power law model. Figure \textbf{11} indicates that power law
model is stable in case of accelerating era as it will decay in future.
\begin{figure}
\centering \epsfig{file=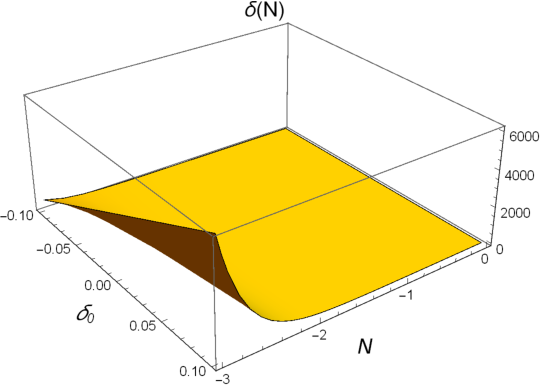, width=.45\linewidth,
height=2in} \textbf{(a)} \epsfig{file=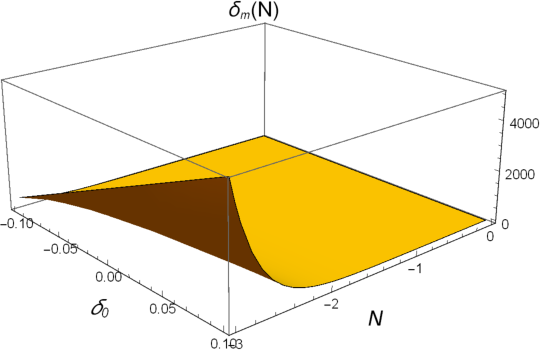, width=.45\linewidth,
height=2in}
\centering \epsfig{file=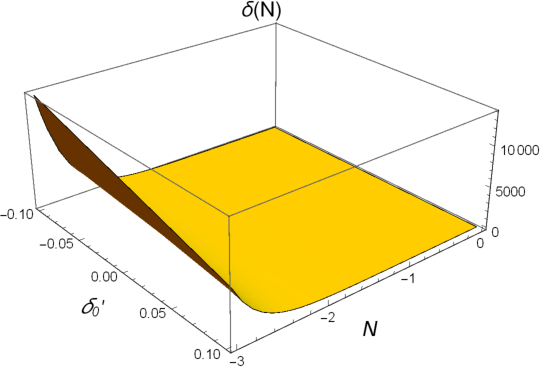, width=.45\linewidth,
height=2in} \textbf{(b)} \epsfig{file=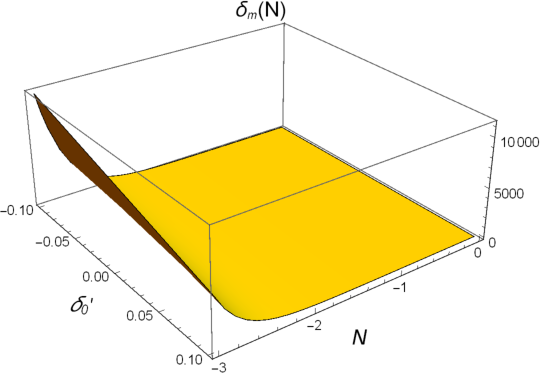, width=.45\linewidth,
height=2in}
\centering \epsfig{file=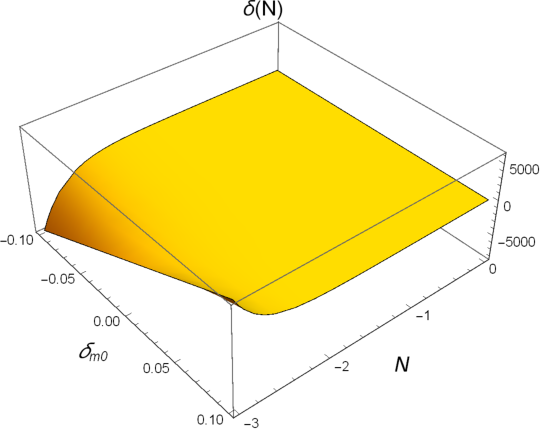, width=.45\linewidth,
height=2in} \textbf{(c)} \epsfig{file=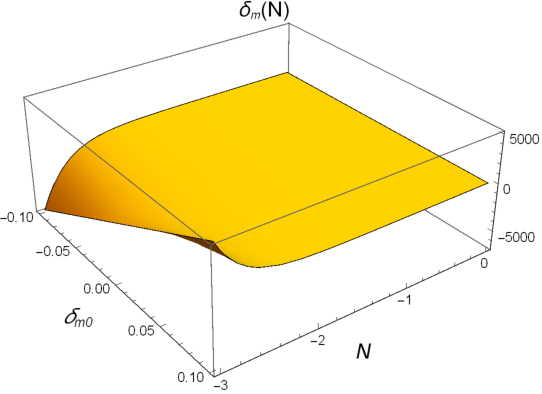, width=.45\linewidth,
height=2in}\caption{\label{figpower} Stability of power law reconstructed model with $\alpha=\frac{2}{3}$
(dominated era). In figure $\delta(N)$ and $\delta_m(N)$ represent the perturbation parameters with initial
conditions on $\delta(N)$, $\delta'(N)$ and $\delta_m(N)$ at $N=0$. In (a), we set $\delta'_0$=$\delta_{m0}=.1$
and vary $\delta_0$, (b) vary $\delta_0'$ and in (c) vary $\delta_{m0}$. Here chosen parameters are
$a_i=1,i=1,2,3,4$, $\lambda_1=10,~\lambda_2=2,~\lambda_3=5$ and $\omega=\frac{2}{3}$.}
\end{figure}
\begin{figure}
\centering \epsfig{file=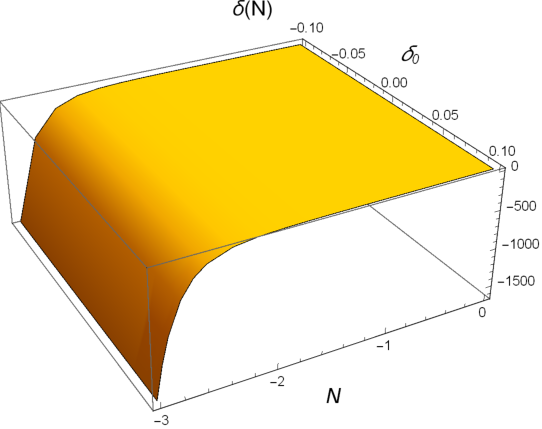, width=.45\linewidth,
height=2in} \textbf{(a)} \epsfig{file=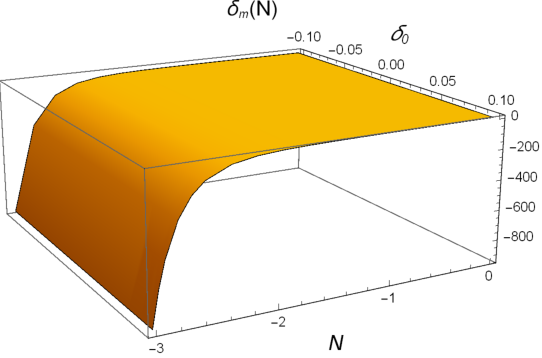, width=.45\linewidth,
height=2in}
\centering \epsfig{file=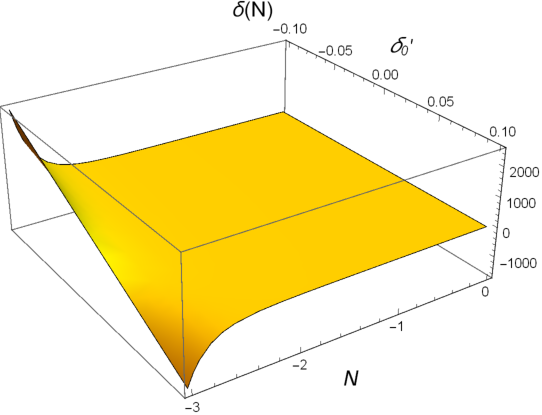, width=.45\linewidth,
height=2in} \textbf{(b)} \epsfig{file=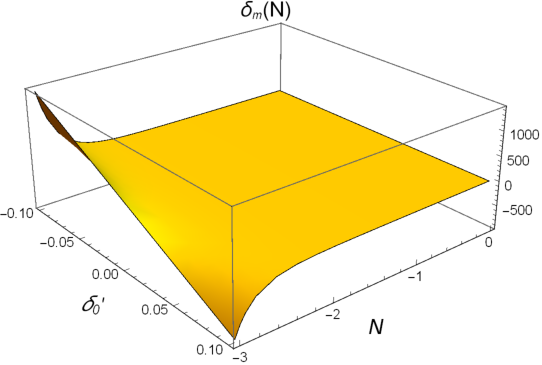, width=.45\linewidth,
height=2in}
\centering \epsfig{file=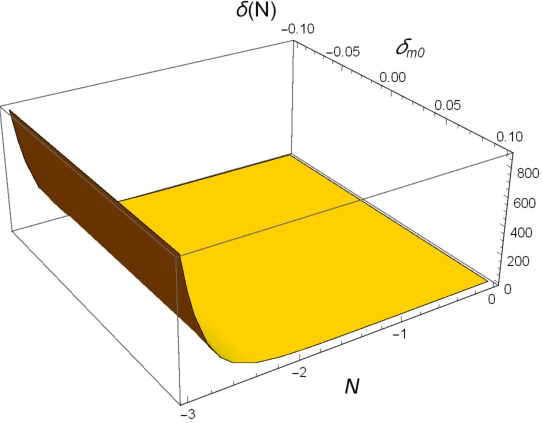, width=.45\linewidth,
height=2in} \textbf{(c)} \epsfig{file=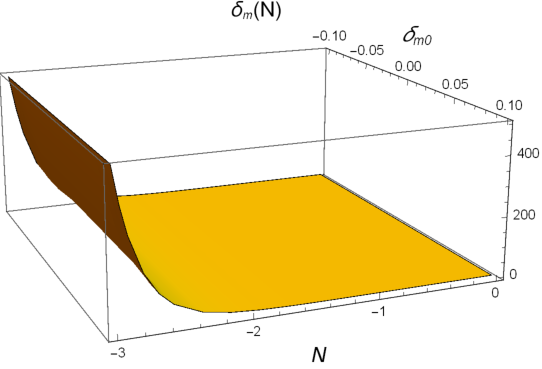, width=.45\linewidth,
height=2in}\caption{\label{figpower1} Stability of power law reconstructed model with $\alpha=2$ (accelerating era).
In figure $\delta(N)$ and $\delta_m(N)$ represent the perturbation parameters with initial conditions on $\delta(N)$,
$\delta'(N)$ and $\delta_m(N)$ at $N=0$. In (a), we set $\delta'_0$=$\delta_{m0}=.1$ and vary $\delta_0$, (b)
vary $\delta_0'$ and in (c) vary $\delta_{m0}$. Here chosen parameters are $a_i=1,i=1,2,3,4$, $\lambda_1=10,~\lambda_2=2,~\lambda_3=5$
and $\omega=\frac{2}{3}$.}
\end{figure}

\subsection{Stability of $\Lambda$CDM Model}

In this segment, we shall examine the stability of reconstructed solution of $\Lambda$CDM model given
in Eq.(\ref{L5}). We can solve the system of perturbed equations numerically for this solution using same
initial conditions as discussed earlier. We present the plots of perturbations in Figure \textbf{12}. The figure
shows that the perturbation parameters $\delta$ and $\delta_m$ do not converge to zero even in future so model
becomes unstable.
\begin{figure}
\centering \epsfig{file=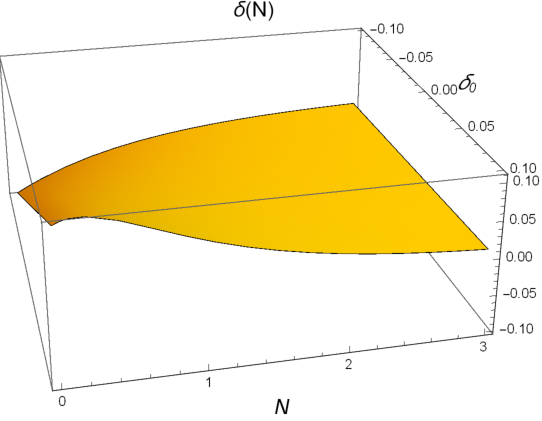, width=.45\linewidth,
height=2in} \textbf{(a)} \epsfig{file=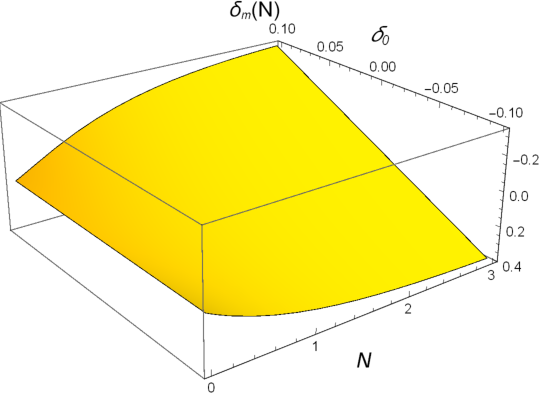, width=.45\linewidth,
height=2in}
\centering \epsfig{file=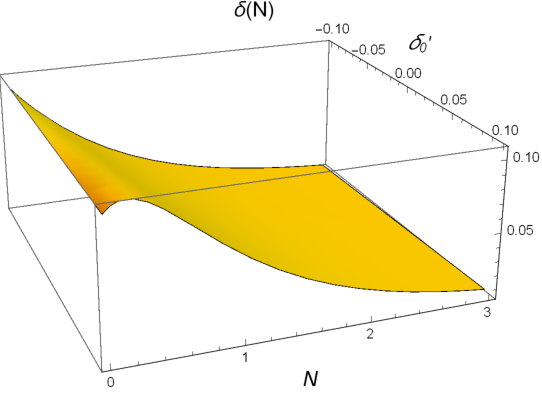, width=.45\linewidth,
height=2in} \textbf{(b)} \epsfig{file=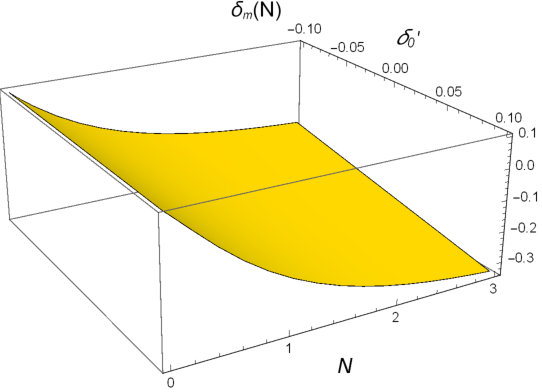, width=.45\linewidth,
height=2in}
\centering \epsfig{file=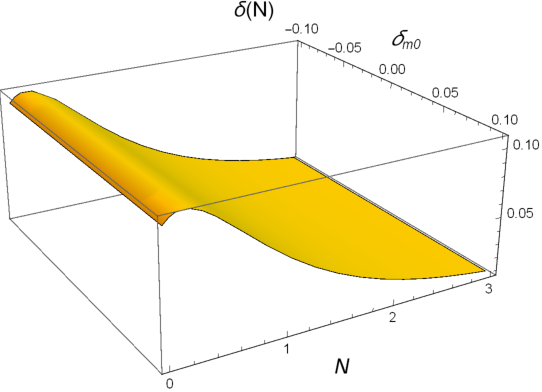, width=.45\linewidth,
height=2in} \textbf{(c)} \epsfig{file=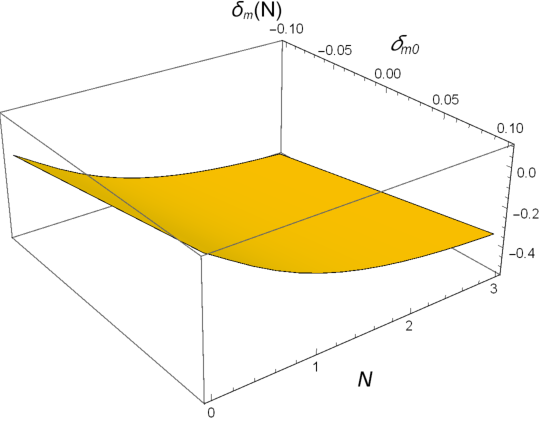, width=.45\linewidth,
height=2in}\caption{\label{figL} Stability of $\Lambda$CDM reconstructed model \ref{L5}. In figure $\delta(N)$
and $\delta_m(N)$ represent the perturbation parameters with different initial conditions. Here chosen parameters
are $a_1=a_2=b_1=b_2=1$, $\lambda_1=10,~\lambda_2=2,~\lambda_3=5$ and $\omega=\frac{2}{3}$.}
\end{figure}

\subsection{Stability of de Sitter model}

Here, we shall analyze the stability of reconstructed de Sitter universe model given in Eq.(\ref{d5}).
Here, we solve the Eqs.(\ref{s8}) and (\ref{s11}) numerically for model (\ref{d5}). For this, we set
$\lambda_1=10$, $\lambda_2=5$ and $\lambda_3=10$ and show the evolution of perturbations $\delta$ and $\delta_m$
by applying initial conditions. To obtain some significant results, we have varied the initial conditions on
$\delta,~\delta^{'}$ and $\delta_m$. The growth of perturbed parameters corresponding to each initial condition
is presented in Figures \textbf{13(a)-(c)}. It is clear that these perturbations incline to zero.
This figure represents that no oscillation occur and it converges to zero shows stable behavior.
\begin{figure}
\centering \epsfig{file=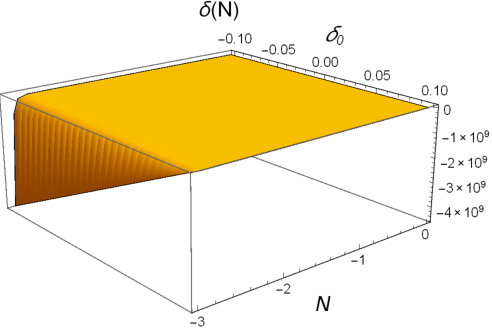, width=.45\linewidth,
height=2in} \textbf{(a)} \epsfig{file=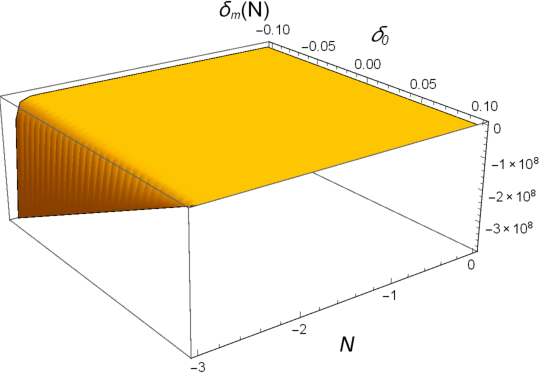, width=.45\linewidth,
height=2in}
\centering \epsfig{file=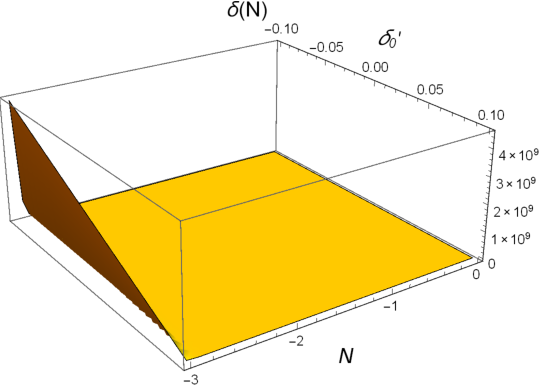, width=.45\linewidth,
height=2in} \textbf{(b)} \epsfig{file=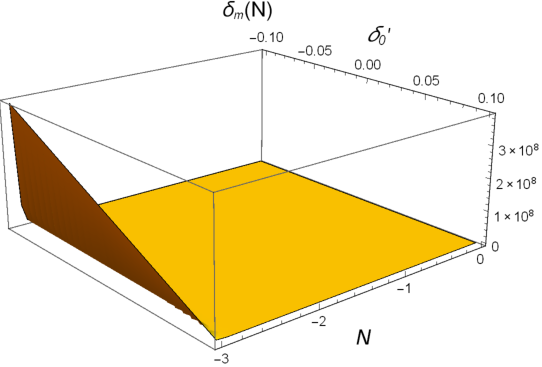, width=.45\linewidth,
height=2in}
\centering \epsfig{file=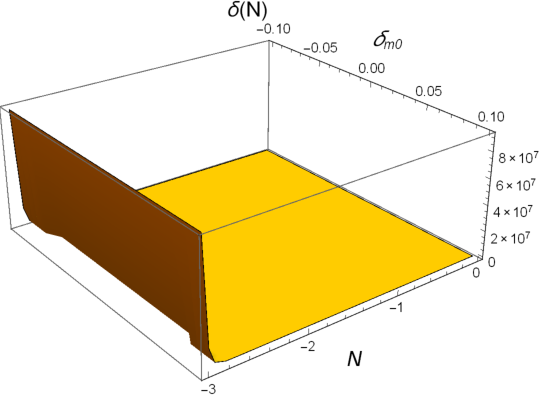, width=.45\linewidth,
height=2in} \textbf{(c)} \epsfig{file=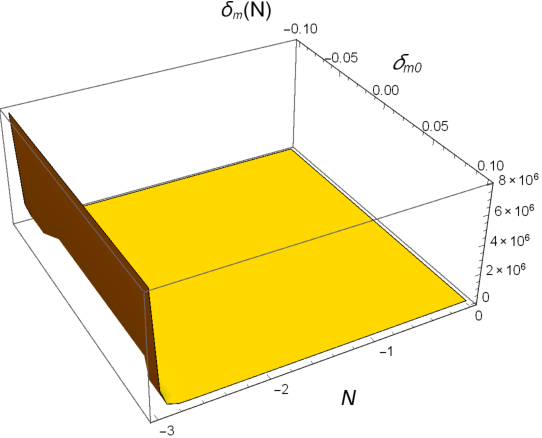, width=.45\linewidth,
height=2in}\caption{\label{figD} Stability of de sitter reconstructed model \ref{d5}.
In figure $\delta(N)$ and $\delta_m(N)$ represent the perturbation parameters with different
initial conditions. Here chosen parameters are $a_1=a_2=c_1=c_2=1$, $\lambda_1=10,~\lambda_2=5,~\lambda_3=10$
and $\omega=\frac{2}{3}$.}
\end{figure}

\section{Concluding Remarks}

$\mathcal{F}(R,L_{m},T)$ gravity unifies two classes of gravitational theories
with geometry-matter coupling, namely the $\mathcal{F}(R,L_{m})$ and the $\mathcal{F}(R, T)$ theories. It generalizes the role of geometry-matter coupling with coupling being introduced between geometry and trace of EMT as well as with the matter Lagrangian $L_m$. Classical theory of relativity is based on simple algebraic structure and its action is formulated as the sum of independent geometrical and physical contents. This theory generalizes the algebraic structure and brings new degrees of freedom into modified dynamical equations. This theory assists in presenting the acceptable cosmic evolution form high to low redshifts as compared to the results of the $\mathcal{F}(R,L_{m})$ and the $\mathcal{F}(R, T)$ theories. One can explore different cosmic aspects and explore the viability of this extended version of non-minimally coupled modified theories. The present manuscript is a significant contribution in this respect where we have studied some interesting cosmological features of this newly proposed gravitational framework.

In this manuscript non-equilibrium description of thermodynamics is presented by defining the Wald entropy relation \cite{W} which results in modified first law of thermodynamics, involving the contribution from entropy production term. Moreover, the role of generalized matter geometry coupling can be seen in the relation of GSLT and one can retrieve the thermodynamics results of well known theories including $\mathcal{F}(R)$, $\mathcal{F}(R,L_{m})$ and the $\mathcal{F}(R, T)$ theories \cite{10m, R10, R11, 39, 40, 42, 44}. Further we have tested the validity of the GSLT by assuming that the overall temperature
within the horizon $T_{tot}$ is equivalent to the temperature of the AH, i.e., ($T_{Ah}$). It is worthy
to mention here that in the present work, for discussing GSLT and other features, we have considered the
conserved EMT by imposing the relevant constraint. Next we have formulated the energy condition bounds in this framework and investigated the validity of WEC by taking three simple models of $\mathcal{F}$ function. In the same segment, we have explored the possibility of getting phantom/quintessence behavior of DE EoS parameter by constraining the involved free parameters. Lastly, we have proposed some new reconstructed $\mathcal{F}(R,T,L_m)$ functions corresponding to some well-known and simple cosmology like de Sitter, power law and $\Lambda$CDM.

The whole discussion presented in this work can be summarized as follows.

\begin{itemize}
\item Firstly, we have explored the validity regions of GSLT constraint by considering three specific
forms of $\mathcal{F}(R,T,L_m)$ function $\mathcal{F}=R+\gamma TL_m+\sigma$, $\mathcal{F}=R+\eta TL_m$
and $\mathcal{F}=R+\psi T+2\zeta L_m$ which are already available in literature \cite{22m}.
The main focus was to find the possible ranges of free $\mathcal{F}(R,T,L_m)$
model parameters so that GSLT remains valid. For this purpose, we generated 3D region graphs of the
corresponding constraints by varying the basic model parameters and explore their possible validity ranges.
Then by using these ranges, we have shown 2D validity region graphs for GSLT in the respective subsections.
\item Energy constraints are developed for this modified framework in general and then the validity of resulting
inequalities for NEC and WEC is investigated. It is worthy to point out here that NEC is trivially valid for all three cases.
In section \textbf{IV}, we have determined the possible ranges of model parameters in detail for which the
WEC remains valid in this theory.
\item In literature, it is a well-known fact that our universe is undergoing phase of accelerated expansion (phantom
or quintessence and hence referring to $\omega_d=-1.073^{+0.090}_{-0.089}$ as per observational data set of
WMAP + eCMB + BAO + H0 \cite{LR}. In the same section \textbf{IV}, we have discussed the behavior of DE EoS parameter
for this framework and determined the possible ranges of free model parameters for which quintessence or phantom
cosmic phases or $\omega_d=-1$ can be attained.
\end{itemize}
\textbf{Some newly proposed models using reconstruction technique:}
\begin{itemize}
\item Next we have adopted the well-known reconstruction methodology to reconstruct the
forms of $\mathcal{F}(R,L_m,T)$ function by taking power law, $\Lambda$CDM and de Sitter cases
into account. Generally it is difficult to get exact solutions due to non-linearity of the resulting
system of equations. In the present work, we have used separable sum form of
$\mathcal{F}=\lambda_1F_1(R)+\lambda_2\mathcal{G}(L_m)+\lambda_3\mathcal{H}(T)$ where we fixed $\mathcal{G}(L_m)=\rho$
for simplicity reasons and reconstructed the form of other functions. It is worthy to mention here that for
all cases, the exact solution was possible.
\item Next we have examined the GSLT constraint for these reconstructed solutions and provided
the corresponding validity regions for different ranges of coupling parameters $\lambda_i$, $i=1,2,3$.
It is shown that this law is valid for power law, $\Lambda$CDM and de sitter models for different values
of coupling parameters. The detail of validity regions for different values of $\lambda_i$ is provided in
Table. \ref{table0}.
\item In the corresponding subsections of Section \textbf{V}, we have shown that the WEC can be
valid for the reconstructed functions if some specific choices of free model parameters are considered.
We also investigated the behavior of DE EoS parameters for these reconstructed functions and discussed
the possibilities to achieve the DE dominated cosmic eras.
\item Lastly, we analyzed the stability of reconstructed solutions of power law, $\Lambda$CDM and de Sitter
models by introducing some suitable perturbations. It is concluded that power law and de Sitter models
are stable while $\Lambda$CDM showed unstable behavior. This behavior is consistent with the results
obtained in \cite{Sum1}, where the $\mathcal{F}(R,T)$ models were reconstructed for the same cases and
found to be stable.
\end{itemize}
Dolgov–Kawasaki instability \cite{Sum2} is another significant criterion to check the stability of
modified theories involving higher-order dynamical equations. In literature \cite{Sum3}, these
conditions have been explored for $\mathcal{F}(R)$ theory with non-minimal interaction of matter and geometry and
the well-known $\mathcal{F}(R,T)$ gravity. Recently, Dolgov–Kawasaki instability is explored for this newly
proposed $\mathcal{F}(R,T,L_m)$ theory \cite{22m}. It is argued that for $\mathcal{F}(R,T)$ and $\mathcal{F}(R,T,L_m)$
theories, this criteria remains the same as in $\mathcal{F}(R)$ gravity and is given by $$\mathcal{F}_{RR}\geq0,~~ R>R_0,$$
where $R_0$ is the Ricci scalar today. It can be easily checked that for three simple choices of $\mathcal{F}$ functions,
this condition is trivially satisfied. For the reconstructed function in de Sitter case, this condition leads
to
\begin{equation}\nonumber
(9.31323\times10^{-10}C_1+5.94829\times10^{13}C_2)\lambda_1\geq0
\end{equation}
which can be easily satisfied when $C_1,~~ C_2,~~\lambda_1>0$ or $C_1,~~ C_2,~~\lambda_1<0$. Here the
same choices of other involved parameters have been used as we picked in graphical analysis.
For the reconstructed power law function, it yields the expression
\begin{equation}\nonumber
\frac{(8a_1+3a_2R^{5/2})\lambda_1}{4R^3}\geq0.
\end{equation}
It can be easily checked that the above inequality can be satisfied if $a_1,~~ a_2,~~\lambda_1>0$ or $a_1,~~a_2,~~\lambda_1<0$.
Lastly, for the reconstructed $\Lambda$CDM model, this criterion takes the following form:
\begin{equation}\nonumber
(\frac{4.06217\times10^{-9}b_1}{(1+7.53982e^{-3N})^{5/2}})\lambda_1\geq0
\end{equation}
where we have assumed $b_2=0$. It can be easily verified that this inequality holds when
$b_1,~~\lambda_1>0$ or $b_1,~~\lambda_1<0$.\\
It would be worthwhile to explore other cosmological aspects of this theory. Also, the proposed reconstructed
models can be more investigated by using phase space analysis and their significance can be revealed.
\begin{table}
\centering \caption{Validity region of GSLT for different ranges of $\lambda_i$} \label{table0}
\begin{tabular}{|c|c|c|c|}
\hline
Models & $\lambda_1$ & $\lambda_2$ & $\lambda_3$  \\\hline
Power law  &$\lambda_1\geq0$  & $\lambda_2=0$ & $\lambda_3\geq 0$\\
                       &$\lambda_1\leq 0$& $\lambda_2=0$ & $\lambda_3\leq -74.67$\\
											& $\lambda_1\leq 0$ & $\lambda_2\leq-50.27$& $\lambda_3=0$\\
											&$\lambda_1\geq 0$ & $\lambda_2\geq-50.26$& $\lambda_3=0$\\
											\hline
$\Lambda$CDM          &$\lambda_1\leq 0$ & $\lambda_2=0$ & $-100\leq\lambda_3\leq 100$\\
                      &$\lambda_1\geq 0$ & $\lambda_2=0$ & $\lambda_3\leq 0$\\
											& $\lambda_1\leq 0$ & $\lambda_2\geq-50$& $\lambda_3=0$\\
											&$\lambda_1\geq 0$ & $\lambda_2\leq-50.5$& $\lambda_3=0$\\
											\hline
de sitter           &$\lambda_1\leq 0$ &$\lambda_2=0$ & $\lambda_3\leq 0$\\
                       &$\lambda_1\geq 0$& $\lambda_2=0$ & $\lambda_3\geq 0$\\
											& $\lambda_1\leq 0$ & $\lambda_2\leq-51$& $\lambda_3=0$\\
											&$\lambda_1\geq 0$ & $\lambda_2\geq-50$& $\lambda_3=0$\\
											\hline
\end{tabular}
\end{table}

\section{Appendix}

\begin{eqnarray}\nonumber
b_0&=&-6H_{\ast}^2F_{R}^{\ast}-(12\rho_{\ast}(H_{\ast}H_{\ast}^{'}+4H_{\ast}^2)
+3(-1+3\omega)H_{\ast}^2\rho_{\ast}^{'})F_{RT}^{\ast}+18(2H_{\ast}^3H_{\ast}^{''}+H_{\ast}^2H_{\ast}^{'^2}
+7H_{\ast}^3H_{\ast}^{'}-4H_{\ast}^4)F_{RR}^{\ast}\\\nonumber
&-&108(H_{\ast}^3H_{\ast}^{''}+H_{\ast}^2H_{\ast}^{'^2} + 4H_{\ast}^3H_{\ast}^{'})(H_{\ast}H_{\ast}^{'}+4H_{\ast}^2)F_{RRR}^{\ast}
+18(-1+3\omega)(H_{\ast}^3H_{\ast}^{'}+4H_{\ast}^4)\rho_{\ast}^{'}F_{RRT}^{\star}\\\nonumber
&&-18(-1+3\omega)(H_{\ast}^3H_{\ast}^{'}+4H_{\ast}^4)\rho_{\ast}^{'}F_{RRL}^{\ast}+(-6\rho_{\ast}(H_{\ast}H_{\ast}^{'}+4H_{\ast}^2)
+3(-1+3\omega)H_{\ast}^2\rho_{\ast}^{'})F_{RL}^{\ast},\\\nonumber
b_1&=&-12\rho_{\ast}H_{\ast}^2F_{R T}^{\ast}+(36H_{\ast}^3 H_{\ast}^{'}+54H_{\ast}^4)
F_{RR}^{\ast}-108H_{\ast}^3(H_{\ast}^2 H_{\ast}^{''}+H_{\ast}H_{\ast}^{'^2}
+4H_{\ast}^2H_{\ast}^{'})F_{RRR}^{\ast}+ 18(-1+3\omega)H_{\ast}^4\rho_{\ast}^{'}F_{RRT}^{\star}\\\nonumber
&&-6H_{\ast}^2\rho_{\ast}F_{RL}^{\ast}-18H_{\ast}^4\rho{\ast}^{'}F_{RRL}^{\ast},\\\nonumber
b_2&=& 18 H_{\ast}^{4}F_{RR}^{\ast},\\\nonumber
c_{m1}&=&8\pi\rho_{\ast}+\frac{(5-\omega)\rho_{\ast}F_{T}^{\ast}}{2}+2(-1+3\omega)\rho_{\ast}^2F_{TT}^{\ast}+
(-1+3\omega)(\rho_{\ast}(3H_{\ast}H_{\ast}^{'}+H_{\ast}^{2})- 3H_{\ast}^2\rho_{\ast}^{'})F_{RT}^{\ast}
+18(-1+3\omega)\rho_{\ast}\\\nonumber
&\times&(H_{\ast}^3H_{\ast}^{''}+H_{\ast}^2H_{\ast}^{'^2}+4H_{\ast}^3H_{\ast}^{'})F_{RRT}^{\ast}-3(-1+3\omega)^2H_{\ast}^2
\rho_{\ast}\rho_{\ast}^{'}F_{RTT}^{\ast}-\rho_{\ast}^2F_{LT}^{\ast}+\frac{\rho_{\ast}}{2}F_{L}^{\ast}
-\rho_{\ast}^2F_{LL}^{\ast}+3H_{\ast}\rho_{\ast}F_{LRR}^{\ast}\\\nonumber
&&-(-1+3\omega)(\rho_{\ast}(3H_{\ast}H_{\ast}^{'}+H_{\ast}^{2})-
3H_{\ast}^2\rho_{\ast}^{'})F_{RL}^{\ast}+3H_{\ast}^2(-1+3\omega)^2\rho_{\ast}\rho_{\ast}^{'}(F_{LRT}^{\ast}
+F_{TRL}^{\ast})-3(-1+3\omega)^2\rho_{\ast}\rho_{\ast}^{'}F_{LRL}^{\ast},\\\nonumber
c_{m2}&=&3(-1+3\omega)H_{\ast}^{2}(-F_{RT}^{\ast}+F_{RL}^{\ast}),\\\nonumber
d_{1}&=&\rho_{\ast}H_{\ast}\{8\pi+\frac{F_{L}^{\ast}}{2}-\frac{(-1+3\omega)F_{T}^{\ast}}{2}
+\rho_{\ast}^2F_{LL}^{\ast}+6(1+\omega)(-1+3\omega)\rho_{\ast} H_{\ast}F_{TT}^{\ast}\},\\\nonumber
d_2&=&3H_{\ast}(1+\omega)\rho_{\ast}^2F_{LL}^{\ast}+\frac{9}{2}H_{\ast}(1+\omega)(1-\omega)\rho_{\ast}F_{T}^{\ast}
-\frac{3}{2}H_{\ast}(1+\omega)(-1+3\omega)\rho_{\ast}^2f_{TT}^{\ast}
-6(1+\omega)(-1+3\omega)^2\rho_{\ast}^2\rho_{\ast}^{'}H_{\ast}f_{TTT}^{\ast}\\\nonumber
&&-3H_{\ast}(1+\omega)(-1+3\omega)\rho_{\ast}^3F_{TLL}^{\ast}+3H_{\ast}(1+\omega)\rho_{\ast}^3F_{LLL}^{\ast},\\\nonumber
d_3&=&(3(\omega-3)H_{\ast}-18(1+\omega)\rho_{\ast}H_{\ast}^2)F_{RT}^{\ast}-6(1+\omega)(-1+3\omega)
\rho_{\ast}\rho_{\ast}^{'}(H_{\ast}^3H_{\ast}^{'}+4H_{\ast}^{4})F_{RTT}^{\ast}
+18\rho_{\ast}^2H^3_{\ast}(1+\omega)F_{RLL}^{\ast},\\\nonumber
d_4&=&3(1+\omega)\rho_{\star}H_{\star}(k^2+F_{T}^{\star})-\{3(3-\omega)\rho_{\star}^{'}+18(1+\omega)\rho_{\star}\}
(H_{\ast}^2H_{\ast}^{'}+4H_{\ast}^{3})F_{RT}^{\ast}-6(1+\omega)(-1+3\omega)\\\nonumber
&\times& \rho_{\ast}\rho_{\ast}^{'}(H_{\ast}^2H_{\ast}^{'}+4H_{\ast}^{3})F_{RTT}^{\ast}-6(1+\omega)(-1+3\omega)\rho_{\ast}
\rho_{\ast}^{'}(H_{\ast}^2H_{\ast}^{'}+4H_{\ast}^{3})F_{RLL}^{\ast}+\frac{3}{2}(1+\omega)H_{\ast}\rho_{\ast}F_{L}^{\ast},\\\nonumber
\hat{b_0}&=&-6H_{\ast}^2F_{R}^{\ast}+18(2H_{\ast}^3H_{\ast}^{''}+ H_{\ast}^2H_{\ast}^{'^2}
+7H_{\ast}^3H_{\ast}^{'}-4H_{\ast}^4)F_{RR}^{\ast}-108(H_{\ast}^3H_{\ast}^{''}+H_{\ast}^2H_{\ast}^{'^2}
+ 4H_{\ast}^3H_{\ast}^{'})(H_{\ast}H_{\ast}^{'}+4H_{\ast}^2)F_{RRR}^{\ast},\\\nonumber
\hat{ b_1}&=&(36H_{\ast}^3 H_{\ast}^{'}+54H_{\ast}^4)
F_{RR}^{\star}-108H_{\ast}^3(H_{\ast}^2 H_{\ast}^{''}+H_{\ast}H_{\ast}^{'^2}
+4H_{\ast}^2H_{\ast}^{'})F_{RRR}^{\ast},\\\nonumber \hat{b_2}&=& 18 H_{\ast}^{4}F_{RR}^{\ast},\\\nonumber
\hat{c_{m1}}&=&-\bigg[8\pi\rho_{\star}+\frac{(5-\omega)\rho_{\ast}F_{T}^{\ast}}{2}+2(-1+3\omega)
\rho_{\ast}^2F_{TT}^{\ast}+\frac{\rho_{\ast}}{2}F_{L}^{\ast}-\rho_{\ast}^2F_{LL}^{\ast}\bigg],\\\nonumber
\hat{ d_{1}}&=& \rho_{\ast}H_{\ast}\{8\pi+\frac{(1-\omega)F_{T}^{\ast}}{2}
+\frac{\rho_{\ast}}{2}F_{L}^{\ast}+\rho_{\ast}F_{LL}^{\ast}+3(1+\omega)(-1+3\omega)\rho_{\ast} H_{\ast}F_{TT}^{\ast}\},\\\nonumber
\hat{d_2}&=& (-1+3\omega)\rho_{\ast}\{(\frac{1}{2}(5+\omega)\rho_{\ast}^{'}H_{\ast}+3(1+\omega)\rho_{\ast}H_{\ast})F_{TT}^{\ast}
+(1+\omega)(-1+3\omega)\rho_{\ast}\rho_{\ast}^{'}H_{\ast}F_{TTT}^{\ast}\}
+3H_{\ast}(1+\omega)\rho_{\ast}^2F_{LL}^{\ast}+\\\nonumber
&&3H_{\ast}(1+\omega)\rho_{\ast}^3F_{LLL}^{\ast},\\\nonumber
\hat{d_4}&=&3(1+\omega)\rho_{\ast}H_{\ast}(k^2+F_{T}^{\ast})+\frac{3}{2}(1+\omega)H_{\ast}\rho_{\ast}F_{L}^{\ast},
\end{eqnarray}

\end{document}